\pdfoutput=1
\documentclass[useAMS,usenatbib]{mn2e}
\pdfoutput=1
\usepackage{enumerate}
\usepackage{amsmath}
\usepackage{graphicx}
\usepackage{color}
\usepackage{hyperref}
 
 \providecommand{\adsurl}[1]{\href{#1}{ADS}}
  
\usepackage{afterpage}
\usepackage{amsfonts}

\usepackage{pifont}

\usepackage{ulem}
\usepackage{array}  
\title{Parameter Estimation with BEAMS in the presence of biases and correlations}
\newcommand{\btp}[0]{$\tau_A$-probability}
\newcommand{\btps}[0]{$\tau_A$-probabilities}
\newcommand{\cfed}[0]{confirmed}
\newcommand{\bs}[1]{\boldsymbol{#1}}

\newcommand{\KBH}[0]{KBH}
\newcommand{\ori}[0]{KBH}

\newcommand{\full}[0]{full}
\author[Newling et al.]{J. Newling$^{1,2}$\thanks{E-mail:
james.newling@gmail.com},  B. Bassett$^{1,2,3}$, R. Hlozek$^{4}$,  M. Kunz$^{5}$, M. Smith$^{2,6}$ , M. Varughese$^{7}$\\
$^{1}$Department of Mathematics and Applied Mathematics, University of Cape Town, Rondebosch 7701, South Africa\\
$^{2}$African Institute for Mathematical Sciences, 6-8 Melrose Road, Muizenberg,  7945, South Africa\\
$^{3}$South African Astronomical Observatory, PO Box 9, Observatory 7935, South Africa\\
$^{4}$Department of Astrophysics, Oxford University, Oxford OX1 3RH, United Kingdom\\
$^{5}$D\'{e}partement de Physique Th\'{e}orique, Universit\'{e} de Gen\`{e}ve, Gen\`{e}ve CH1211, Switzerland\\
$^{6}$Astrophysics, Cosmology and Gravity Centre, University of Cape Town, Rondebosch 7701, South Africa\\
$^{7}$Department of Statistical Sciences, University of Cape Town, Rondebosch 7701, South Africa}

\begin{document}

\date{Not submitted to arXiv at last compile: \today}

\pagerange{\pageref{firstpage}--19} \pubyear{2010}
\maketitle

\label{firstpage}

\begin{abstract}

The original formulation of BEAMS - Bayesian Estimation Applied to Multiple Species - showed how to use a dataset contaminated by points of multiple underlying types to perform unbiased parameter estimation. An example is cosmological parameter estimation from a photometric supernova sample contaminated by unknown Type Ibc and II supernovae. Where other methods require data cuts to increase purity, BEAMS uses all of the data points in conjunction with their probabilities of being each type. Here we extend the BEAMS formalism to allow for correlations between the data and the type probabilities of the objects as can occur in realistic cases. We show with simple simulations that this extension can be crucial, providing a 50\% reduction in parameter estimation variance when such correlations do exist. We then go on to perform tests to quantify the importance of the type probabilities, one of which illustrates the effect of biasing the probabilities in various ways. Finally, a general presentation of the selection bias problem is given, and discussed in the context of future photometric supernova surveys and BEAMS, which lead to specific recommendations for future supernova surveys.

\end{abstract}


\section{Introduction}

Type Ia Supernovae (SNeIa) provided the first widely accepted evidence for cosmic acceleration in the late 1990s \citep{sn1,sn2}. While they were based on relatively small numbers of spectroscopically-confirmed SNeIa, those results have since been confirmed by independent analyses of other data sets\citep{bao1,bao2,clusters,lensing,ISW,bao3,CMB}.

Next generation SN surveys such as LSST will be fundamentally different, yielding thousands of high-quality candidates every night for which spectroscopic confirmation will probably be impossible. Creating optimal ways of using this excellent photometric data is a key challenge in SN cosmology for the coming decade.
There are two ways that one can imagine using photometric candidates. The first approach is to try to classify the candidates into Ia, Ibc or II SNe \citep{jocro, kuzncoll, poodles, fuzzy} and then use only those objects that are believed to be SNeIa above some threshold of confidence. This has recently been discussed by \cite{sako1} who showed that photometric cuts could achieve high purity. Nevertheless it is clear that this approach can still lead to biases and systematic errors from the small contaminating group when used in conjunction with the simplest parameter estimation approaches such as the maximum likelihood method. 

A second approach is to use all the SNe, irrespective of how likely they are to actually be a SNIa. This is the approach exemplified by the BEAMS formalism, which accounts for the contamination from non-Ia SN data using the appropriate Bayesian framework, as presented in~\cite{beams}, hereafter referred to as KBH.
In KBH, two threads are woven: a general statistical framework, and a discussion of how it may be applied to SNeIa. As noted in KBH, the general framework can be applied to any parameter estimation problem involving several populations, and indeed may have already been done so in other fields. In this paper we take the same approach as in KBH of keeping the notation general enough for application to other problems, while discussing its relevance to SNe.

We will attempt to use the same notation as in~\KBH{}, but differ where we consider it necessary. For example, we write conditional probability functions as 
$f_{\Theta|D}(\theta|d)$.
The quantity
$f_{\Theta|D}(\theta|d) \Delta\theta,$
should be interpreted as the probability that $\Theta$ lies in the interval $(\theta, \theta + \Delta\theta)$, conditional on $D = d$ (for small $\Delta \theta$).

We preserve capital letters for random variables and lowercase letters for their observed values.  In the BEAMS framework, one wishes to estimate parameter(s) $\Theta$ from $N$ observations of the random variable $X$. We will use the boldface $\bs{X}$ to denote a vector of $N$ such random variables: $\bs{X} = X_{1 \cdots N}$. An observation of $X$ we will denote by $x$, so that the full set of $N$ observations is denoted by $\bs{x} = x_{1 \cdots N}$. For SNe, the observations $\bs{x}$ are the photometric data of the $N$ SNe. As such, for SNe the probability density function (pdf) $f_{\bs{X}|\Theta}(\bs{x}|\theta)$ is the likelihood of observing the photometric data $\bs{x}$ assuming some cosmological parameters  $\theta$, which we will discuss. The relationship between raw photometric data $(X)$ and the true cosmological parameters $(\Theta)$ is highly intricate, resulting in a pdf which cannot realistically be worked with, and so one first reduces each observation $x$ to a single feature $d$ for which there is a direct $\Theta$-dependent model. For SNe, if the parameters $\Theta$ are for example $\Omega_{\Lambda}$ and $\Omega_m$, then $d$ will consist of an estimated luminosity distance and redshift. If the parameter of interest $\Theta$ is a luminosity distance at a given redshift, then $d$ will be simply a fitted distance modulus. Unless stated otherwise, this is the case.

The correct treatment of redshifts will be important to BEAMS as applied to future SN surveys. Future surveys will likely have only photometric information for the SNe but will have a spectroscopic redshift for the host galaxy obtained by chance (because of overlap with existing surveys) or through a targeted follow-up program. The SDSS-II supernova survey~\citep{DATA_RELEASE1} is an example of both of these. There were host redshifts available from the main SDSS galaxy sample and there was also a targeted host followup program as part of the BOSS survey. Future large galaxy surveys like SKA, EUCLID or BigBOSS will likely provide a very large number of host galaxy redshifts for free.

BEAMS is unique in that the underlying types of the observations are not assumed known. In the case where there are two underlying types ($ T \in \{A, B\}$), each observation has an associated type probability $(P)$ of being type $A$,
\begin{equation*}
 P \stackrel{\text{def}}{=} \textrm{P} (T = A | X_{P}),
\end{equation*}
where $X_{P}$ is a subset of features of $X$. In other words, $X_P$ is the component of the raw data $X$ on which type probabilities are conditional. Note that we treat $P$ as a random variable: while the value of $P$ is completely determined by $X_P$, which in turn is completely determined by $X$, $X$ is a random variable and therefore so too is $P$. The realizations of the type probabilities $\bs{P}$ of the $N$ observations are denoted by $\boldsymbol{p} = p_{1 \cdots N}$, and we will call them \btps{}. The \btp{} for a SN is thus the probability of being type Ia, conditional on knowing the subset $x_P$ of the the photometric data. $x_P$ may be the full photometric time-series, the earliest segment of the SN's light curve, a fitted shape parameter, or any other extracted photometric information.  

Finally, we mention that the type of the SN $(T)$ is a random variable with realisation denoted $\tau$. A summary of all the variables used in the paper is given in Table~\ref{table}.

Attempts to approximate \btps{} include those of \cite{Pozn1, newlingetal, INCA} and as implemented in SALT2~\citep{2007AA46611G}. Note that values obtained using these methods are only approximations of \btps{}, as the algorithms are trained on only a handful of spectroscopically confirmed SNe. Note too that there is no sense in which one set of \btps{} is \textit{the} correct set, as this depends on what $X_P$ is. 
Obtaining unbiased estimates of \btps{} is not easy, and we will consider the problems faced in doing so in Section~\ref{obtubp}. For SNe, the problem is made especially difficult by the fact that spectroscopically confirmed SNe, which are used to train \btp{} estimating algorithms, are brighter than unconfirmed photometric SNe.

In 2009 the Supernova Photometric Classification Challenge (SNPCC) was run to encourage work on SN classification by lightcurves alone~\citep{sntc}. Performance of the classification algorithms was judged according to the final purity and efficiency of extracted Ia samples. While the processing of photometric data is essential to the workings of BEAMS for SNe, the classification of objects is not required. It would be interesting to hold another competition where entrants are required to calculate~\btps{} for SNe. Algorithms would then not only need to recognise SNeIa, but would also need to provide precise, unbiased probabilities of the object being an SNeIa.

In brief, this paper consists of three more or less independent parts. In Section~\ref{impa} we present an extension of BEAMS to the case where certain correlations, which were ignored in \KBH{}, are present. In Section~\ref{gpriors}, we discuss the relevance of \btps{} in a broader context, and specifically the importance to of them in BEAMS. Then is Sections~\ref{sec:sims}, \ref{basicbias} and~\ref{dpcorr}, we perform simulations to better understand the importance of sample sizes, nearness of population distributions, biases of \btps{} and decisivenesses of \btps{} (to be defined). Finally, in Section~\ref{obtubp} we present new ideas from the machine learning literature describing when and how \btp{} biases emerge and how to correct for them. This is then discussed in the context of the SNPCC in Section~\ref{snphotcc}.

\begin{table}
\begin{center}
\label{table}

\begin{tabular}{|c|c|m{5.8cm}|}
\hline
\multicolumn{3}{|c|}{Random Variables} \\
\hline\hline
R.V.  & Data & Definition \\
\hline
$P$  & $p$ & The probability of being type $A$ conditional on $X_P$. We call $P$ the \btp{}. \\
\hline
$D$  & $d$ & A particular feature of an object whose distribution depends directly on the parameter(s) we wish to approximate using BEAMS. SNe: $D$ is luminosity distance.\\
\hline
$T$  & $\tau$ & The type of an object, $T \in \{A, B\}$ SNe: $T \in \{ \text{Ia, nIa} \}$\\
\hline
$X$ & $x$ & All the features observed of an object. SNe: $X$ is the photometric data. \\
\hline
$X_F$ & $x_F$ & That part of the features which affects confirmation probability. SNe: $X_F$ are peak apparent magnitudes. \\
\hline
$X_P$ & $x_P$ & That part of the features used to determine the \btp{}. SNe: $X_P$ could be any reduction of $X$. \\
\hline
$F$ & $f$ & Whether the object is \cfed{} or not. For SNe: $F = 1$ if a spectroscopic confirmation is performed.\\
\hline
$\bar{P}$ & $\bar{p}$ & Is exactly $P$ if the object is unconfirmed and 1 or 0 if \cfed{}, depending on type.\\
\hline
\end{tabular}
\end{center}
\caption{A description of all the random variables used in this paper.}
\end{table}

\section{Introducing and Modifying the Beams equations}
\label{impa}
The posterior probability on the parameter(s) $\Theta$, given the data $\boldsymbol{D}$, is derived in Section II of~\KBH{} as 

\begin{align}
f_{\Theta | \bs{D}}(\theta |\bs{d} ) \propto f_{\Theta}(\theta)   \hspace{3mm} \times &  \label{BEAMSWE} \\
 \sum_{\bs{\tau} \in \left[ A,B \right]^N} & f_{\bs{D} | \Theta, \bs{T}} (\bs{d} |\theta,  \bs{\tau}) \prod_{\tau_i = A}p_i\prod_{\tau_j = B}(1 - p_j),
\notag
\end{align}
where the $p_i$s are \btps{}. The summation is over all of the $2^N$ possible ways that the $N$ observations can be classified into two classes. We will refer to the expression on the right of~\eqref{BEAMSWE} as the \emph{\ori{}} posterior. When the $N$ observations are assumed to be independent, that is when 
\begin{equation*}
f_{\bs{D}|\Theta, \bs{T}}(\bs{d}|\theta, \bs{\tau}) = \prod_{i = 1}^{N} f_{D_i | \Theta, T_i}(d_i | \theta, \tau_i),
\end{equation*}
the \ori{} posterior reduces,
\begin{equation}
\prod_{i = 1}^{N} \left[ \, f_{D_i|\Theta, T_i }\left(d_i| \theta, A \right) p_i +
 f_{D_i|\Theta  ,  T_i}(d_i | \theta,  B) \left(1-p_i\right) \right].  \label{BEAMSIND}
\end{equation}

There is one substitution in the derivation of the \ori{} posterior on which we would like to focus, given in \KBH{} as eqn. (5) on page 3:

\begin{equation}
\label{wrong}
f_{\bs{T}}(\bs{\tau}) = \prod_{\tau_i = A}p_i \prod_{\tau_i = B}\left(1-p_i\right).
\end{equation}

Equation~\eqref{wrong} states that the l.h.s. prior probability of the SNe having types $\bs{\tau}$ is given by the product on the r.h.s. involving \btps{}. We argue that this product should not be treated as the prior $f_{\bs{T}}$, but rather as the conditional $f_{\bs{T} | \bs{P}}$. In effect, we argue that KBH should not use the \btps{} $\bs{p}$ unless $\bs{P}$ is explicitly included as a conditional parameter. It is to this end that we now rederive the posterior on $\Theta$, taking $f_{\Theta | \bs{D}, \bs{P}}(\theta |\bs{d}, \bs{p})$ as a starting point, discussing at each line what has been used.

\begin{align} 
& \hspace{-2mm}f_{\Theta |  \bs{D}, \bs{P}}(\theta | \bs{d}, \bs{p})  &   \notag 
\intertext{$\rightarrow$ We will first use the definiton of conditional probability to obtain,}\notag
& = \frac{f_{\Theta, \bs{D}, \bs{P}}(\theta, \bs{d}, \bs{p})}{f_{\bs{D}, \bs{P}}(\bs{d}, \bs{p})}. \notag \intertext{$\rightarrow$ The term in the numerator can then be written as the sum over all $2^N$ possible type vectors,}\notag
 &  = \sum_{\boldsymbol{\tau}} \frac{f_{\Theta, \bs{D}, \bs{P}, \bs{T}}(\theta, \bs{d}, \bs{p}, \bs{\tau})}{f_{\bs{D}, \bs{P}}(\bs{d}, \bs{p})}. \notag 
\intertext{$\rightarrow$ The numerator is again modified using the definition of conditional probability,} \notag
 &  = \sum_{\boldsymbol{\tau}} \frac{f_{\bs{D} | \Theta, \bs{P}, \bs{T}}(\bs{d}| \theta, \bs{p}, \bs{\tau})f_{\bs{\Theta}, \bs{P}, \bs{T}}(\theta, \bs{p}, \bs{\tau})}{f_{\bs{D}, \bs{P}}(\bs{d}, \bs{p})}.\notag 
\intertext{$\rightarrow$ We will now assume that the probability of having \btps{} and types $\bs{p}$ and $\bs{\tau}$ respectively are independent of $\Theta$. As noted following eqn.(4) in \KBH{}, for SNe this assumption rests on the fact that $\Theta$ (that is $\Omega_{m}$, $\Omega_{\Lambda}$) describes large scale evolution, while the SN types $\bs{\tau}$ depend on local gastrophysics, with little or no dependence on perturbations in dark matter.}\notag 
 &  = \sum_{\boldsymbol{\tau}} \frac{f_{\bs{D} | \Theta, \bs{P}, \bs{T}}(\bs{d}| \theta, \bs{p}, \bs{\tau})f_{\Theta}(\theta) f_{\bs{P}, \bs{T}}(\bs{p}, \bs{\tau})}{f_{\bs{D}, \bs{P}}(\bs{d}, \bs{p})}. \notag 
 \intertext{$\rightarrow$ Rearranging this, and again using the definition of conditional probability, we obtain,} \nonumber
 &  = \frac{f_{\bs{P}}(\boldsymbol{p})}{f_{\bs{D}, \bs{P}}(\boldsymbol{d}, \boldsymbol{p})} f_{\Theta}(\theta)   \sum_{\boldsymbol{\tau}}  f_{\bs{D} | \Theta, \bs{P}, \bs{T}}(\bs{d} | \theta, \bs{p}, \bs{\tau})f_{\bs{T} | \bs{P}}(\boldsymbol{\tau}|\bs{p}). \nonumber 
\intertext{$\rightarrow$ The first term on the line above is constant with respect to $\Theta$, and so is absorbed into a proportionality constant. We now make one final weak assumption: $f_{\bs{T}|\bs{P}}(\bs{\tau}|\bs{p}) = \prod_{i = 1}^N f_{T_i | P_i} (\tau_i | p_i)$. This assumption will be necessary to make a comparison with the \ori{} posterior. Making this assumption we arrive at,} \notag
   &  \propto f_{\Theta}(\theta) \sum_{\boldsymbol{\tau}}  f_{\bs{D} | \Theta, \bs{P}, \bs{T}}(\bs{d} |  \theta, \bs{p}, \bs{\tau}) \prod_{\tau_i = A}p_i\prod_{\tau_j = B}(1 - p_j). \label{pp} \\
\end{align}
We will refer to the newly derived expression~\eqref{pp} as the \full{} posterior. Let us now consider the difference between the \ori{} posterior~\eqref{BEAMSWE} and the \full{} posterior, and notice that in the \full{} posterior, the likelihood of the data $\bs{D}$ is conditional on $\Theta, \bs{P}$ and $\bs{T}$, whereas in the \ori{} posterior $\bs{D}$ is only conditional on $\Theta$ and $\bs{T}$. This is the only difference between the two posteriors, and so when $\boldsymbol{D} \vert \Theta, \boldsymbol{T}$ is independent of $\boldsymbol{P}$, the posterior~\eqref{pp} reduces to the \ori{} posterior \eqref{BEAMSWE}, making them equivalent. This is an important result: when $\boldsymbol{D} \vert \Theta, \boldsymbol{T}$ and $\bs{P}$ are independent, the \ori{} and \full{} posteriors are the same. 

\medskip

\noindent Our results can be summarised as follows,

\begin{align} 
\intertext{(1)  As the posterior $f_{\Theta | \bs{D}}(\theta |\bs{d})$ is not conditional on \btps{} it should be independent of \btps{}, and we thus prefer to replace the \ori{} posterior in~\eqref{BEAMSWE} by}\notag
f_{\Theta | \bs{D}}(\theta |\bs{d} ) \propto f_{\Theta}(\theta)   \hspace{3mm} \times &  \notag \\
 \sum_{\bs{\tau} \in \left[ A,B \right]^N} & f_{\bs{D} | \Theta, \bs{T}} (\bs{d} |\theta,  \bs{\tau}) \prod_{\tau_i = A}\pi\prod_{\tau_j = B}(1 - \pi), \notag \\
\intertext{where $\pi$ is an estimate of the global proportion of type $A$ objects. \medskip  \newline (2)   $f_{\Theta | \bs{D}, \bs{P}}(\theta |\bs{d}, \bs{p})$ is always given by the \full{} posterior~\eqref{pp}. When $\boldsymbol{D} \vert \Theta, \boldsymbol{T}$ and $\boldsymbol{P}$ are independent, it reduces to the \ori{} posterior~\eqref{BEAMSWE}.}\notag
\end{align} 
 
It is worth discussing for SNe the statement, ``$\boldsymbol{D} \vert \Theta, \boldsymbol{T}$ and $\boldsymbol{P}$ are not independent''. One \emph{incorrect} interpretation of this statement is, ``given that we know the cosmology is $\Theta$, observing\footnote{Of course we mean ``observing'' in the statistical sense, that is obtaining the realistation of the \btp{} $(p)$ with some software} $\bs{P}$ for a SN of unknown type adds no information to the estimation of the distance modulus.'' Indeed it is difficult to imagine how this could be the case: we know that SNeIa are brighter that other SNe, and therefore obtaining a \btp{} close to 1 shifts the estimated distance modulus downwards (towards being brighter). 
 
A correct interpretation of the statement is, ``given the cosmology $\Theta$, observing $\bs{P}$ of a SN \emph{of known type} adds no information to the estimation of the distance modulus.'' It may seem necessarily true that a \btp{} contributes no new information if the type of the SN is already known, but this is not in general the case; it depends on the method by which \btps{} are obtained.
  
Currently for SNe, fitted distance moduli and approximations of \btps{} are frequently obtained simultaneously, using for example SALT2~\citep{2007AA46611G}. This in itself suggests that $\boldsymbol{D} \vert \Theta, \boldsymbol{T}$ and $\boldsymbol{P}$ will not be independent. In some cases however, \btps{} are calculated from the early stages of the light curves~\citep{Sulli06,Sako07} while the distance modulus is estimated from the peak of the light curve, and so the dependence may be weak. As another example, in Section 4.4 of \cite{newlingetal} \btps{} are obtained directly from a Hubble diagram. Objects lying in regions of high relative SNIa density are given higher \btps{} than objects lying in low relative SNIa density. As a result, at a given redshift, brighter nIa SNe have higher \btps{} than faint nIa SNe. Similarly, at a given fitted distance modulus (fitted assuming type Ia), nIa will lie on average at lower redshifts than Ia. Both of these cases, (distance modulus $|$ $\Theta$, type) being correlated with $P$, and (redshift $|$ $\Theta$, type) being correlated with $P$, are precisely when $\boldsymbol{D} \vert \Theta, \boldsymbol{T}$ and $\boldsymbol{P}$ are dependent. In Section~\ref{dpcorr} a simulation illustrating this dependence is presented.

For completeness, we mention that in the case of independent observations, that is when,
\begin{equation*}
f_{\bs{D}|\Theta, \bs{P}, \bs{T}}(\bs{d}|\theta, \bs{p}, \bs{\tau}) = \prod_{i = 1}^{N} f_{D_i | \Theta, P_i, T_i}(d_i | \theta, p_i, \tau_i),
\end{equation*}
the full posterior~\eqref{pp} reduces to,
\begin{align}
f_{\Theta | \bs{D}, \bs{P}}(\theta | \bs{d}, \bs{p}) \propto \prod_{i = 1}^{N} \left[ \, f_{D_i|\Theta, P_i, T_i}\left(d_i\right | \theta,  p_i, A) p_i \hspace{3mm} + \right. \label{BEAMSINDTRUE}  \\
\hspace{6mm} f_{D_i|\Theta,  P_i, T_i}(d_i|\theta, p_i,  B) & \left(1-  p_i\right)   \big].\notag
\end{align}

\medskip

In Section~\ref{obtubp} we will make suggestions as to what functional form may be chosen for $f_{D_i|\Theta, P_i, T_i}$ when using BEAMS for independent SNe.

\section{Rating \btps{}}
\label{gpriors}
An object's \btp{} is the expected proportion of other objects with its features which are type $A$. In other words, if an object has features $x$, its \btp{} is the expected proportion of objects with features $x$ which are type $A$. Suppose that the global distribution of $P$ is $f_P$. 
The expected total proportion of type $A$ objects is then
\begin{equation}
\label{expp}
\textrm{P}(T = A) = \left< P \right> = \int_{0}^{1} p f_P(p) \, dp.
\end{equation}
In some circumstances, it is necessary to go beyond calculating \btps{} and commit to an absolute classification, as was the case in the SNPCC. In such cases the optimal strategy moving from a \btp{} to an absolute type ($A$ or $B$) is to classify objects positively ($A$) when the \btp{} is above some threshold probability $(c)$. The False Positive Rate (FPR) using such a strategy is
\begin{eqnarray}
\textrm{FPR}(f_P) &=&  \textrm{P}(P>c \vert T = B) \nonumber \\
	&=& \frac{\int_c^1 (1-p)f_P(p) \, dp}{\int_0^1 (1-p)f_P(p) \, dp},\label{fpr}
\end{eqnarray}
and the False Negative Rate is
\begin{eqnarray}
\textrm{FNR}(f_P) &=&  \textrm{P}(P<c \vert T = A) \nonumber \\
	&=& \frac{\int_0^c p \,f_P(p) \, dp}{\int_0^1 p \,f_P(p) \, dp}\label{fnr}.
\end{eqnarray}
 
\begin{figure}
\includegraphics[scale=0.6]{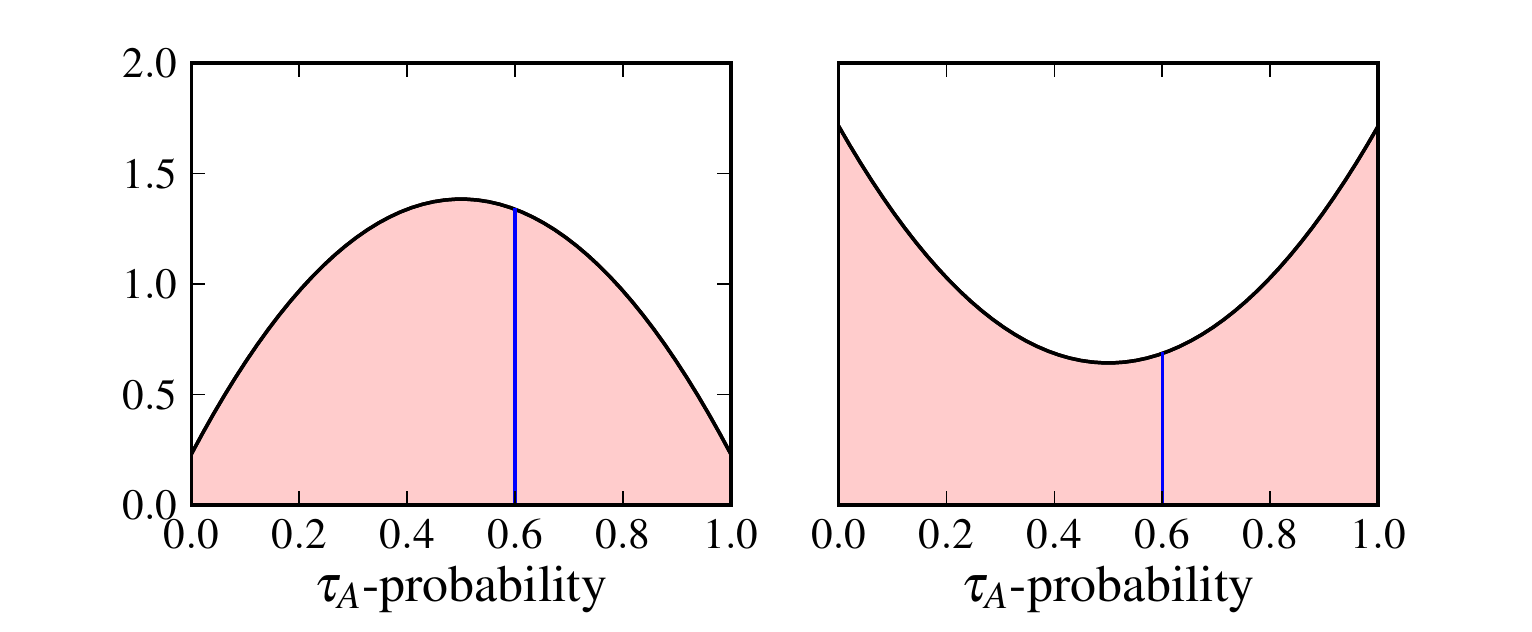} 
\caption{Two \btp{}{} distributions, both with means of $0.5$. Using a threshold of 0.6, we have on left: FPR = 0.17, FNR = 0.45 and on right: FPR: 0.15, FNR~=~0.28}
\label{2distros}
\end{figure}  

For SNe the FPR is the proportion of nIa SNe which are misclassified, while the FNR is the proportion of SNeIa which are misclassified (missed).

\medskip

Intuition dictates that for classification problems, a useful $f_P$ will be one whose mass predominates around 0 and 1. That is, an $f_P$ which with high probability attaches \emph{decisive}\footnote{we say $p_1$ is more decisive than $p_2$ if $|p_1 - 0.5| > |p_2 - 0.5|$.}~\btps{} to observations. To minimize the FPR and FNR this is optimal, as illustrated in Figure~\ref{2distros}. 

We will be presenting a simulation illustrating how the decisiveness of \btps{} affects the parameter estimation of BEAMS. To simplify our study of the effect of the decisiveness of \btps{} on BEAMS, we introduce a family of distributions: For each $\mathcal{P} \in [0.5, 1]$ we have the distribution
\begin{equation}
\label{mathcaP}
f^{\mathcal{P}}(p) = \frac{1}{2} \left( \delta_{\mathcal{P}}(p) + \delta_{1-\mathcal{P}}(p) \right)
\end{equation}
where $\delta_{\mathcal{P}}$ and $\delta_{1- \mathcal{P}}$ are $\delta$-functions centered at $\mathcal{P}$ and $1 - \mathcal{P}$ respectively. It is worth mentioning that we will be drawing probabilities from this distribution, which is potentially confusing. Drawing a observation of $P$ from~\eqref{mathcaP} is equivalent to drawing it from $\left\lbrace 1-\mathcal{P}, \mathcal{P}\right\rbrace$ with equal probability: 
\begin{equation*}
\textrm{P}(P = p)  = \left\{ \begin{array}{ll} 
0.5 & \mbox{if } p =  \mathcal{P} \\
0.5 & \mbox{if } p = 1-\mathcal{P}.
\end{array} \right.
\end{equation*}

If $\mathcal{P}_1$ is more decisive than $\mathcal{P}_2$, we say that the distribution $f^{\mathcal{P}_1}$ is more decisive than $f^{\mathcal{P}_2}$.

On page 5 of~\KBH{} it is stated that the expected proportion of type $A$ objects~\eqref{expp} determines the expected error in estimating a parameter which is independent of population $B$. Specifically, they present the result that the expected error when estimating a parameter $\mu$ with $N$ objects using BEAMS is given by,
\begin{equation}
\label{two}
\sigma_{\mu} \propto \sqrt{\left<P\right> N}.
\end{equation}

It should be noted that the the result from KBH~\eqref{two} is an asymptotic result in $N$. For small $N$, the decisiveness of the probabilities plays an important part. 
If~\eqref{expp} were the only factor determining the expected error ($\sigma_{\mu}$), then $f^{0.5}$ would be equivalent to $f^{1}$ in terms of expected error. This would mean that perfect type knowledge does not reduce error, which would be surprising. An example in Section~\ref{sim2} illustrates that decisiveness does play a role in determining the error. 

As mentioned on page 8 of~\KBH{}, the effect of biases in \btps{} on BEAMS can be catastrophic. Therein they consider the case where there is a uniform bias ($a$) of the \btps{}. That is, if observation $i$ has a claimed \btp{}{} $p_i$ of being type $A$, then there is a real probability $p_i - a$ that it is type $A$. \KBH{} show how, by including a free global shift parameter, such a bias is completely removed. However it is not clear what to do if the form of the bias is unknown. For example, it could be that there is an `overconfidence' bias, where to obtain the true \btps{} one needs to transform the claimed priors ($\tilde{p}$) by 
\begin{equation}
\label{b1}
p = 0.2 + 0.6 \, \tilde{p}.
\end{equation}

Introducing a bias such as the one defined by~\eqref{b1} will have no effect on the optimal FPR and FNR, provided the probability threshold is chosen optimally. This is because~\eqref{b1} is a one-to-one biasing, and so a threshold ($\tilde{c}$) on biased probabilities results in exactly the same partitioning as a threshold in the unbiased space of $0.2 + 0.6\tilde{c}$. However, introducing a bias such as~\eqref{b1} does have an effect on BEAMS parameter estimation, as we show in Section~\ref{basicbias}. In Section~\ref{obtubp} we discuss how to guarantee that the \btps{} are free of bias.

\section{Effects of Decisiveness and sample size on beams}\label{sec:sims}
In this section we will perform simulations to better understand the key factors in BEAMS. The data generated will have the following cosmological analogy: $\Theta$ - distance modulus at a given redshift $z_0$; $\bs{d}$ - the fitted distance moduli of SNe at $z_0$. Furthermore, $\boldsymbol{D} \vert \Theta, \boldsymbol{T}$ and $\boldsymbol{P}$ will be independent, such that the \ori{} and \full{} posterior are equivalent.
 
\subsection{Simulation 1: Estimating a population mean}
\label{sim2}
This simulation was performed to see how the performance of BEAMS is affected by the decisiveness of \btps{}, and by the size of the data set. The two populations ($A$ and $B$) were chosen to have distributions, 
\begin{equation}
\label{moomoo}
f_{D|T}(d,\tau) = \textrm{Normal}(\mu_{\tau}, 1),
\end{equation} 
where $\mu_{A} = -1$ and $\mu_{B} = +1$, as illustrated in Figure~\ref{e1_2d}. The \btp{} distribution is chosen to be $f^{\mathcal{P}}$, so that about half of the observations have a \btp{} of $\mathcal{P}$, with the remaining observations having \btps{} of $1 - \mathcal{P}$. By varying $\mathcal{P}$ we vary the decisiveness.

Let us make it clear how the data for this simulation is generated. First, a \btp{} $(p)$ is selected to be either $\mathcal{P}$ with probability $0.5$ or $1 - \mathcal{P}$ with probability $0.5$, that is according to $f^{\mathcal{P}}$. Second, the type of the observation is chosen, with probability $p$ it is chosen as $A$, and with probability $1 - p$ it is chosen as $B$. Finally, the data $(d)$ is drawn from~\eqref{moomoo}. Notice that $D|T$ is independent of $P$, and so the \ori{} posterior is equivalent to the \full{} posterior. 

In this simulation we only estimate $\mu_A$, with all other parameters known. We use the following Figure of Merit to compare the performance with different sample sizes ($N$) and decisivenesses ($\mathcal{P}$):
\begin{equation*}
h(N, \mathcal{P}) = \frac{1}{ \left<\hat{\mu}_A - \mu_A\right>^2},
\end{equation*}

\begin{figure}
\includegraphics[scale=0.6]{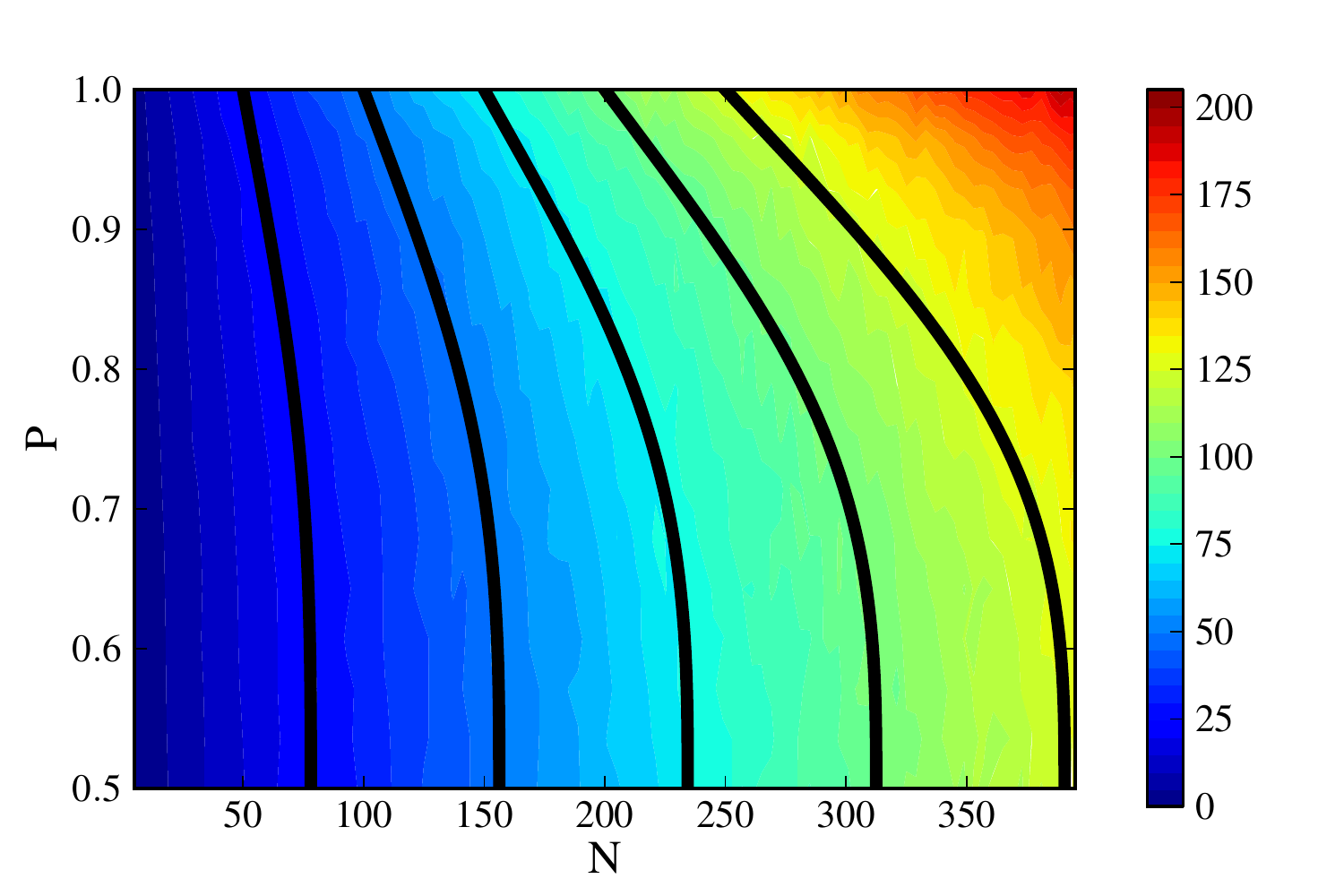} 
\caption{Contour plot of $h(N,\mathcal{P})$. The solid lines are approximations to lines of constant $h$, of the form~\eqref{guess}.} 
\label{cont1}
\end{figure} 

where $\hat{\mu}_A$ is the maximum likelihood estimate of $\mu_A$ using the \ori{} posterior on a sample of size $N$ with \btps{} from $f^{\mathcal{P}}$, and $\left\langle \cdot \right\rangle$ denotes an expectation. Values of $h$ were obtained by simulation, illustrating in Figure~\ref{cont1} the performance of BEAMS for various $(N, \mathcal{P})$ combinations. A good approximation to the FoM $h$ in Figure~\ref{cont1} appears to be
\begin{equation}
\label{guess}
h(N, \mathcal{P}) \approx N \left(0.32 + 1.44(\mathcal{P} - \frac{1}{2})^3\right),
\end{equation}
although this is an ad hoc observation. One interesting observation is that $h(N, \mathcal{P} = 1) \approx h(1.5N, \mathcal{P} = 0.5)$  in the region illustrated in Figure~\ref{cont1}. This says that given a completely blind sample ($\mathcal{P} = 0.5$), and the option to either double its size $(N \rightarrow 2N)$ or to discover the hidden types $(\mathcal{P}: 0.5 \rightarrow 1)$, doubling its size will provide more information about ${\mu}_A$. It is important to reiterate that, according to previously mentioned result of KBH, in the limit of $N \rightarrow \infty$ we do not expect $\mathcal{P}$ to play any part in determining $h(N, \mathcal{P})$. That is, for $N$ sufficiently large, the FoM will be independent of $\mathcal{P}$.

While this simulation is too simple to make extrapolations about cosmological parameter estimations from, it may suggest that the information contained in unconfirmed photometric data may be currently underestimated.

\subsection{Simulation 2: Estimating two population means}
\label{sim22}
\begin{figure}
\includegraphics[scale=0.6]{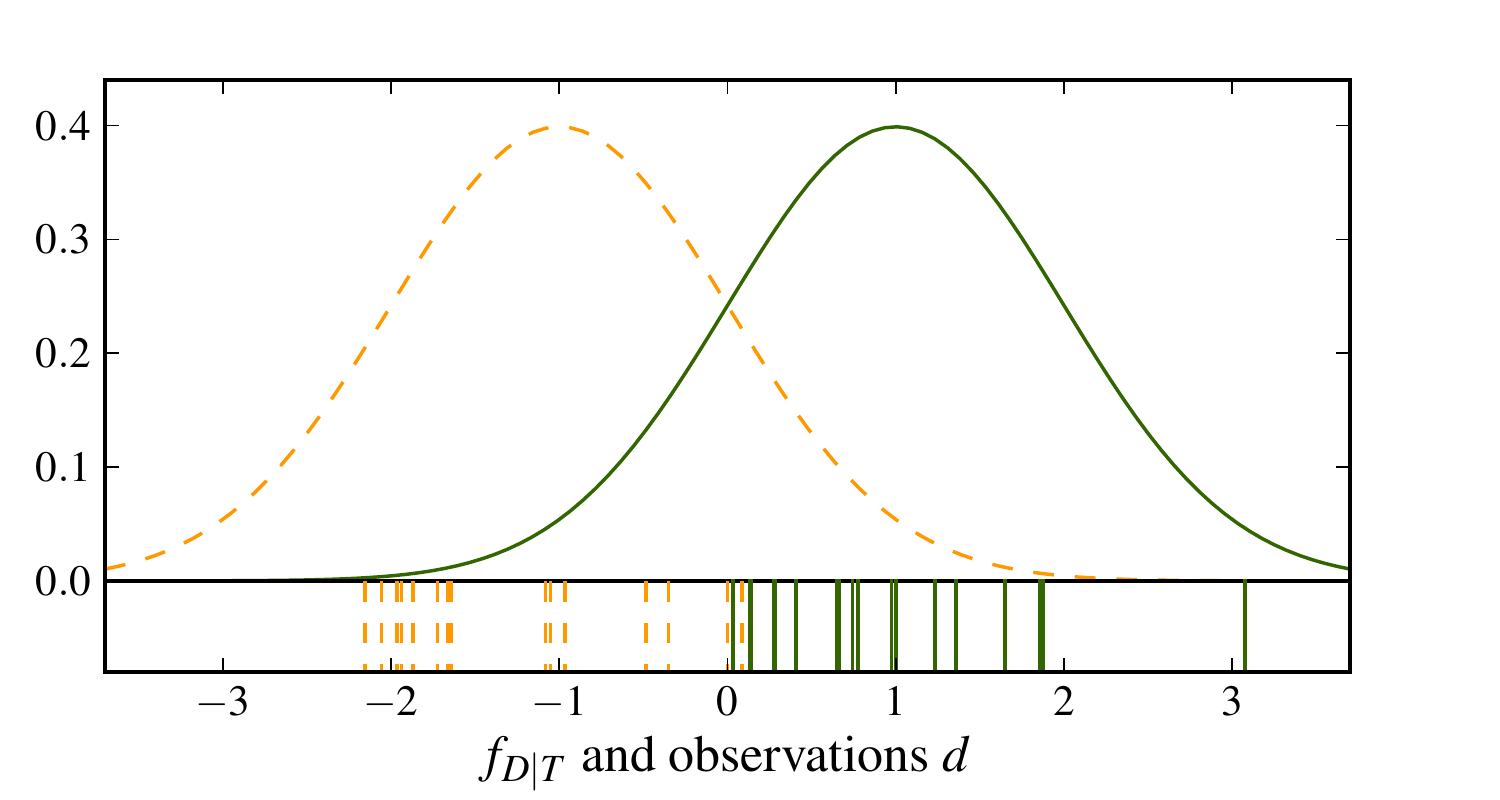} 
\caption{Above are the population $A$ (left) and population $B$ (right) distributions, with (for Simulation \ref{sim22}) the observed values of $D$  drawn from these distributions shown as vertical lines beneath.}
\label{e1_2d}
\end{figure}

\begin{figure}
\includegraphics[scale=0.6]{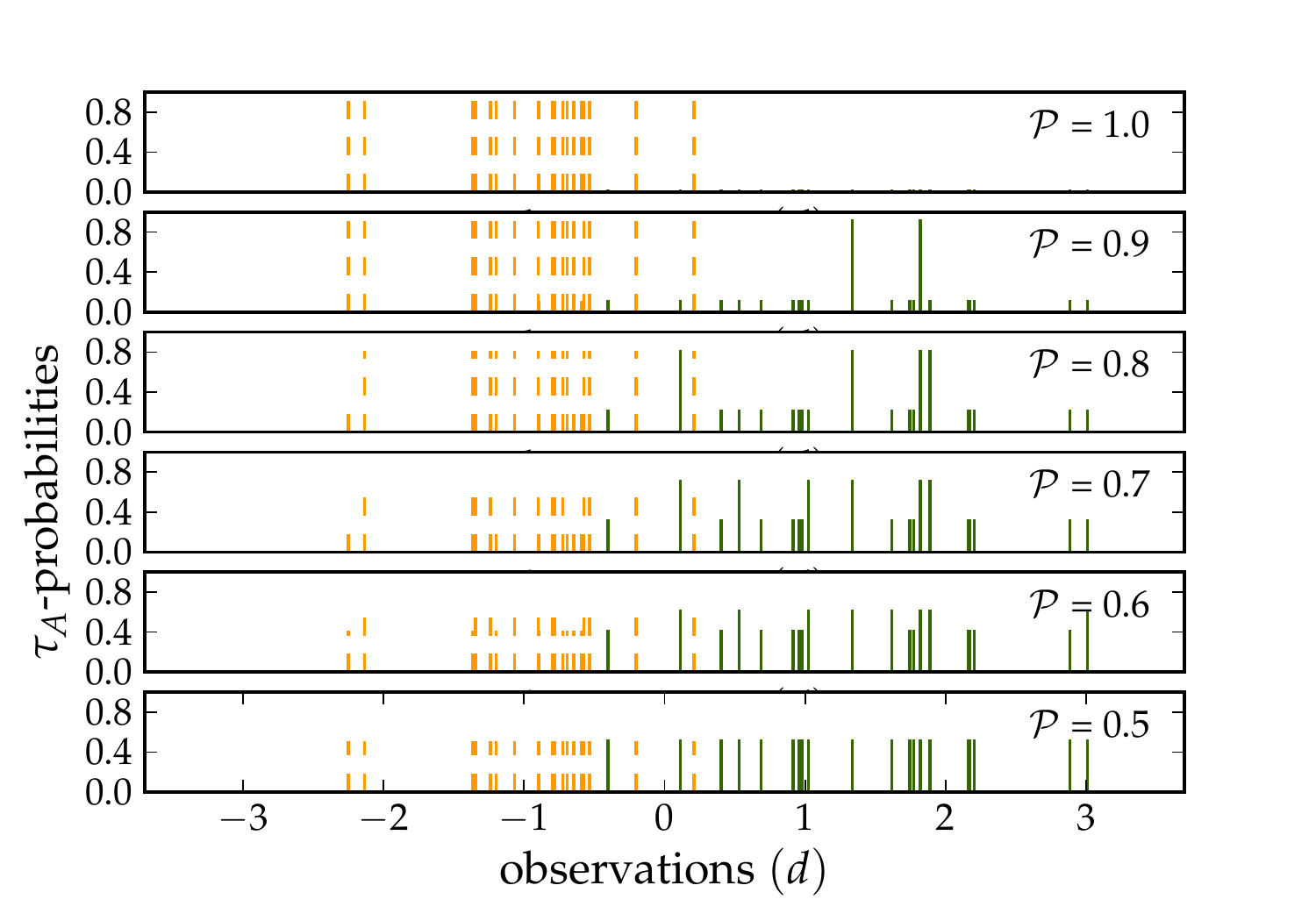} 
\caption{For values of $\mathcal{P}$ from 1 (above) to 0.5 (below), a \btp{} of $\mathcal{P}$ or $1 - \mathcal{P}$ is attached to each observation.}
\label{e1_ws}
\end{figure}

\begin{figure}
\includegraphics[scale=0.6]{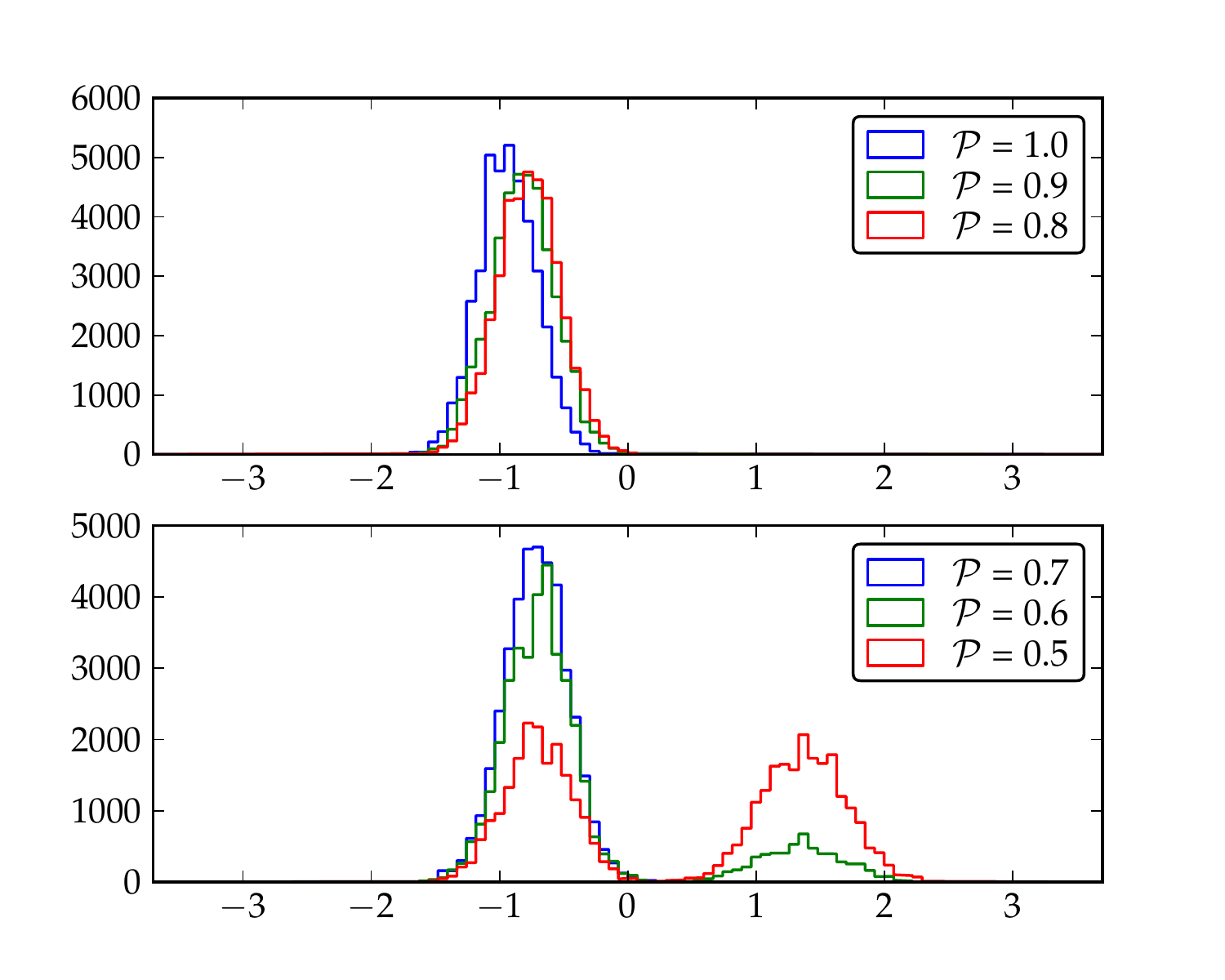} 
\caption{MCMC chain counts, approximating the posterior distributions of $\mu_A$ for the different values of decisiveness, $\mathcal{P}$.}
\label{e1_mar}
\end{figure}

The two population distributions for this simulation are the same as those presented in Simulation 1 and as illustrated in Figure~\ref{e1_2d}. In this simulation, we leave both the population means as free parameters to be fitted for. Twenty objects are drawn from the types $A$ and $B$, with the \btps{} are drawn from $f^{\mathcal{P}}$. The simulation is done with five different $\mathcal{P}$ values. The \btps{} are illustrated in Figure~\ref{e1_ws}, and the approximate shape of the posterior marginals of $\mu_A$ for each $\mathcal{P}$ value are illustrated in Figure~\ref{e1_mar} by MCMC chain counts.   

There are two interesting results from this simulation. The first is that there is negligible difference in performance between $\mathcal{P} = 1$ and $\mathcal{P} = 0.7$, so that having a 30\% type uncertainty for all objects as opposed to absolute type knowledge does not weaken the results. The second is that as $\mathcal{P}$ approaches 0.5, BEAMS still correctly locates the population means but is unsure which mean belong to which population. 

\section{Effects of {\Large{$\tau_A$}}-probabilities bias on BEAMS} 
\label{basicbias}

In the previous section we considered the effect of the decisiveness of \btps{} on the performance of BEAMS. In this section we will consider the effect of using incorrect \btps{}. We will again be estimating $\mu_A$ and $\mu_B$ where they are $-1$ and $1$ respectively, and the population variances are again both known to be $1$. The true \btp{} distribution will be $f^{0.8}$, that is 

\begin{equation*}
\textrm{P}(P = p) = \left\{ \begin{array}{ll} 
0.5 & \mbox{if } p =  0.8 \\
0.5 & \mbox{if } p = 0.2
\end{array} \right.
\end{equation*}

It is worth reminding the reader that we are drawing probabilities from a probability distribution, an unusual thing to do. To generate a \btp{} from this distribution, one could flip a coin, and return $p=0.2$ if $H$ and $p = 0.8$ if $T$.  We consider the effect of biasing \btps{} generated in such a manner in the following ways:

\begin{enumerate}
\item $\mathcal{P} \stackrel{-+}{\Rightarrow} \{0,1 \}$. Here the decisiveness of the \btps{} is overestimated, so that $p = 0.8 \rightarrow p = 1$ and $p = 0.2 \rightarrow p = 0$.\label{no1}

\item $\mathcal{P} \stackrel{+-}{\Rightarrow} \{0.4,0.6\}$. Here the decisiveness of the \btps{} is underestimated, so that $p = 0.8 \rightarrow p = 0.6$ and $p = 0.2 \rightarrow p = 0.4$. \label{no2}

\item $\mathcal{P} \stackrel{--}{\Rightarrow} \{0, 0.6\}$. Here the \btps{} are underestimated by $0.2$, so that $p = 0.8 \rightarrow p = 0.6$ and $p = 0.2 \rightarrow p = 0$.\label{no3}

\item $\mathcal{P} \stackrel{++}{\Rightarrow} \{0.4, 1 \}$. Here the \btps{} are overestimated by 0.2, so that $p = 0.8 \rightarrow p = 1$ and $p = 0.2 \rightarrow p = 0.4$.\label{no4}

\item $\mathcal{P} \stackrel{\sigma}{\Rightarrow} U$. Here, to each \btp{} a uniform random number from $[-0.2, 0.2]$ is independently added.\label{no5}

\end{enumerate} 	

\begin{figure}
\includegraphics[scale=0.6]{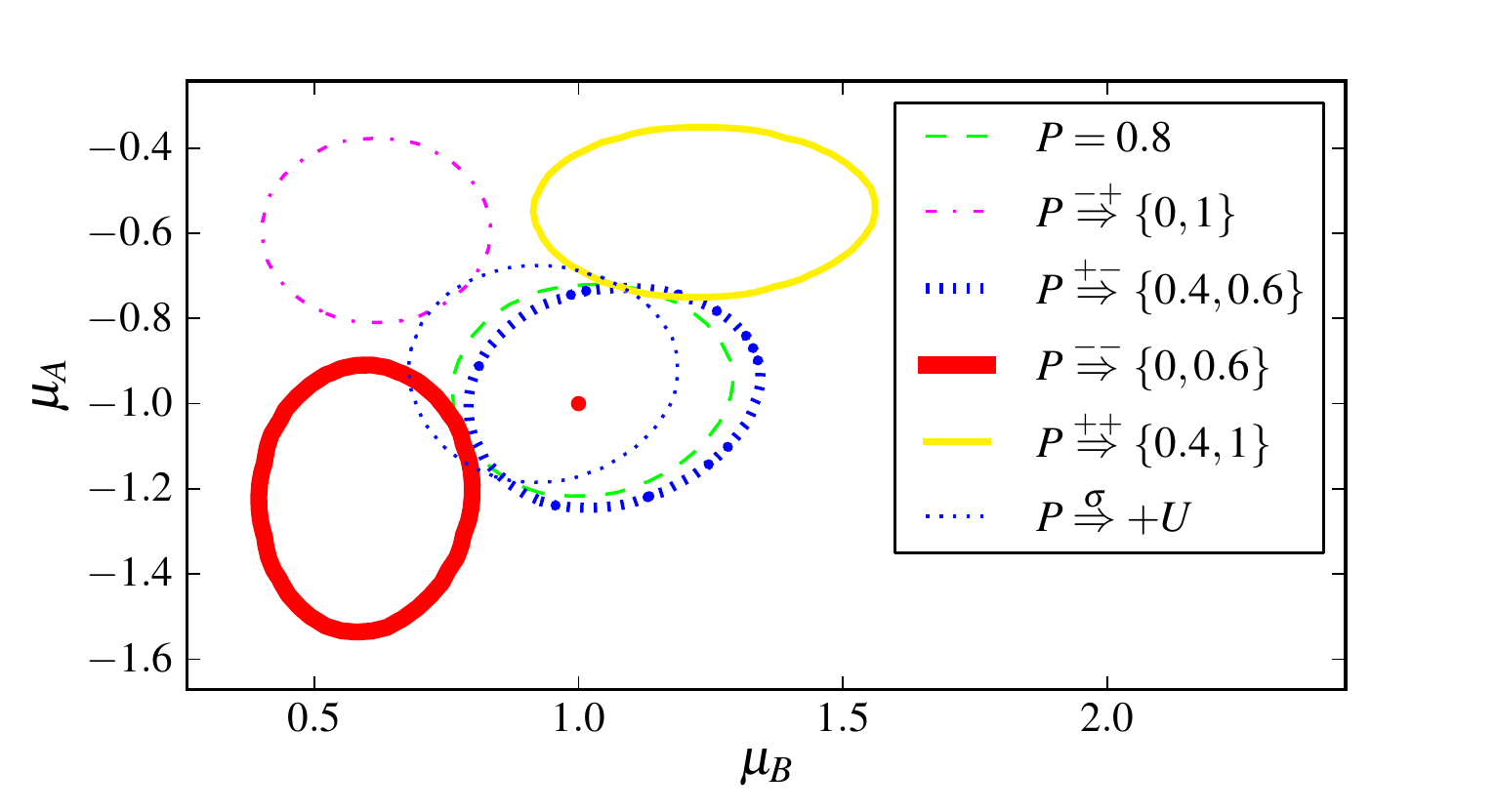} 
\caption{The 99 \% posterior confidence regions using the five biasings of the \btps{}, as described in Section~\ref{basicbias}.}
\label{simp_bias}
\end{figure} 

The 99\% posterior confidence regions obtained using these biased \btps{} in a simulation of 400 points are illustrated in Figure~\eqref{simp_bias}. The underestimation of decisiveness \ref{no2} has little effect on the final confidence region, but overestimating the \btp{} decisiveness \ref{no1} results in a $6\sigma$ bias. Note that overestimating decisiveness results in the estimate $(\hat{\mu}_A, \hat{\mu}_B)$ being biased towards $(\mu_B, \mu_A)$. This is caused by type $B$ objects which are too confidently believed to be type $A$, which pull $\hat{\mu}_A$ towards $\mu_B$, and type $A$ objects which are too confidently believed to be type $B$, which pull $\hat{\mu}_B$ towards $\mu_A$.

The contrast in effect between underestimating and overestimating the decisiveness of \btps{} is interesting, and not easy to explain. One suggestion we have received is to consider the cause of the observed effect as being analogous to the increased contamination rate induced by overestimating the decisiveness in the case BEAMS is not used. With an increased contamination rate comes an increased bias, precisely as observed in Figure~\ref{simp_bias}. It is worth mentioning that underestimating the decisiveness is not entirely without effect, as simulations with more pronounced drops in $\mathcal{P}$ $(0.95 \rightarrow 0.55)$ result in noticable increases in the size of the 99\% confidence region.  

The effect of the flat \btp{} shifts~\ref{no3} and~\ref{no4} introduce biases larger than $4\sigma$. This case was considered in~\KBH{} where, as we have already mentioned, it was shown that simultaneously fitting for this bias completely compensates for it. While this is a pleasing result, one would prefer to know that the \btps{} are correct, as one cannot be sure what form the biasing will take. 

One phenomenon which is observed in this simulation, as it was in simulations as summarised in Table II on page 8 of~\KBH{}, is that a flat \btp{}{} shift in confidence towards being type $B$~\ref{no3} does not bias the estimate of $\mu_A$ as much as it does the estimate of $\mu_B$, and vica versa. In other words, underestimating the probabilities that objects are type $A$ will result in less biased population $A$ parameters than overestimating the probabilities. This result may also be understood in light of an analogy to increased contamination versus reduced population size in the case where BEAMS in not used. 

Finally, we notice that in this simulation the addition of unbiased noise to the \btps{}~\ref{no5} has an insignificant effect. This suggests that systematic biases should be the primary concern of future work on the estimation of \btps{}.
  
\section{When given type, the data is still dependent on {\Large{$\tau$}}-priors}
\label{dpcorr}
\begin{figure}
\includegraphics[scale=0.6]{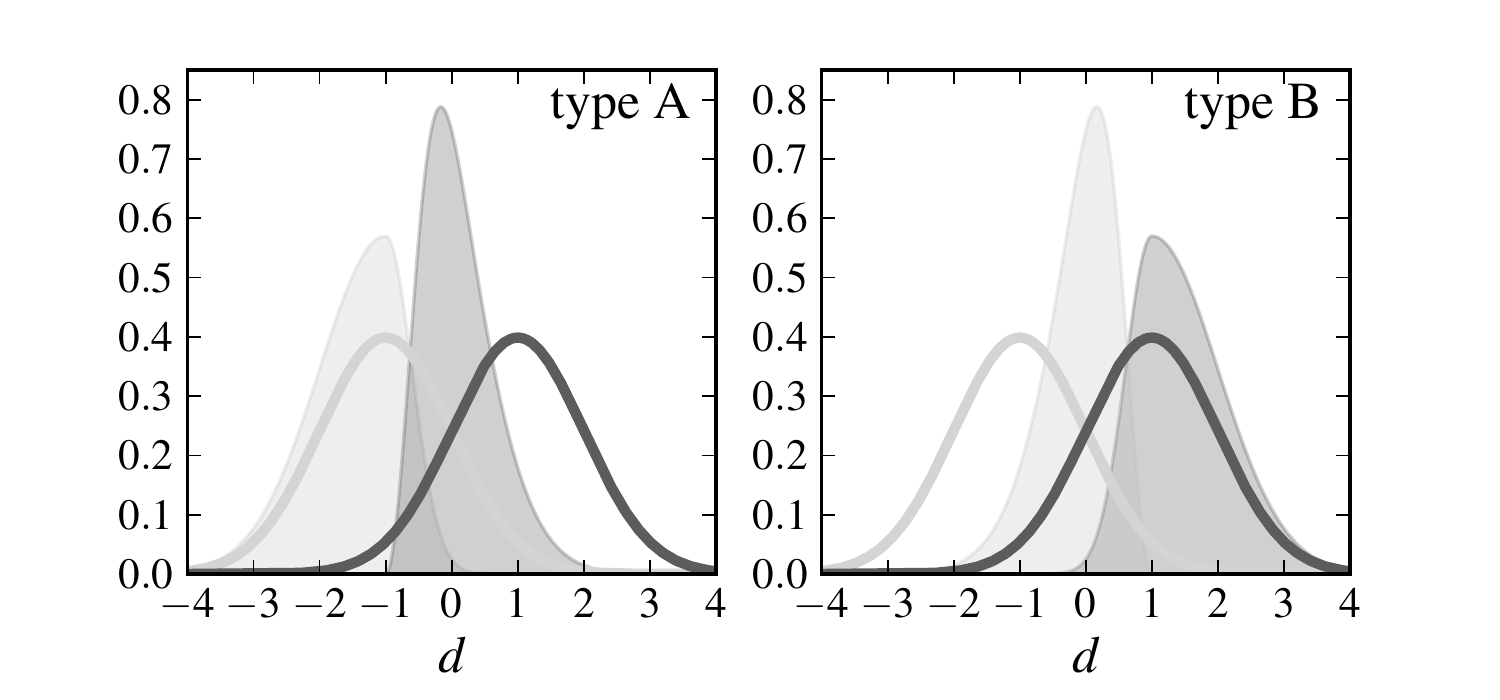} 
\caption{ Plots of $f_{D|P,T}(d|p,\tau)$ (filled curves) for $p$ = 0.7 (light) and $p$ = 0.3 (dark), and for type $A$ (left) and type $B$ (right). Overlying are $f_{D|T}(d|A)$ (light) and $f_{D|T}(d|B)$ (dark).}
\label{pd_corr_dists}
\end{figure}

In this section we consider for the first time a simulation in which the data is not drawn from $f_{D|T}$, but from $f_{D|T,P}$, so that there is a dependence of the data on the \btp{} even when the type is known. The conditional pdfs are shown in Figure~\ref{pd_corr_dists}. To clarify the difference between this simulation and the previous ones, prior to this data was simulated as follows:
\begin{equation*}
P \rightarrow T | P \rightarrow D|T,
\end{equation*}
where at the last step, the data was generated with a dependence only on type. Now it will be simulated as: 
\begin{equation*}
P \rightarrow T | P \rightarrow D|P,T.
\end{equation*}
More specifically, to generate data we start by drawing a \btp{} from $f^{0.7}$,
\begin{align*}
\textrm{P}(P = p) &= \left\{ \begin{array}{ll} 
0.5 & \mbox{if } p =  0.7 \\
0.5 & \mbox{if } p = 0.3.
\end{array} \right. 
\end{align*} 

Note that the above distribution guarantees that $\textrm{P}( T = A) = \frac{1}{2}$. When the \btp{} $(p)$ has been generated, we draw a type ($\tau$) from $\left\{A, B\right\}$ according to 
\begin{align*}
\textrm{P}(T = \tau) &=  \left\{ \begin{array}{ll} 
p & \mbox{if } \tau = A \\
1 - p & \mbox{if } \tau = B.
\end{array} \right.  
\end{align*}

Once we have $p$ and $\tau$, we generate $d$.
The marginals $f_{D|P,T}(d|p, \tau)$ have been chosen such that we have
\begin{align}
f_{D | T}(d|A) &= \text{Normal}\,(-1,1)\label{poo1}\\
f_{D |T}(d|B) &= \text{Normal}\,(1,1),\label{poo2}
\end{align}

as before. The marginal $f_{D|P,T}(d |0.7,A)$ is composed of the halves of two Gaussian curves with different $\sigma$s, chosen such that the tail away from the $B$ population is longer than the one towards the $B$ population. Specifically,  
\begin{align*}
 f_{D|P,T}( d |0.7,A) &=  \hspace{20mm} \\ 
  = &\left\{\begin{array}{ll} 
K \exp \, -\frac{1}{2}(d + 1)^2 & \mbox{if } d<-1 \\
K \exp \, -\frac{100}{32}(d + 1)^2 & \mbox{if } d>-1 \\\end{array} \right. 
\end{align*}

where $K$ is a normalizing constant. The marginal $f_{D|P,T}(d |0.3,A)$ is then constructed to guarantee~\eqref{poo1}. The above construction guarantees that the population of $A$ objects with low \btps{} (0.3) lie on average closer to the $B$ mean than do objects with high (0.7) \btps{}. The marginals of the $B$ population are constructed to mirror exactly the $A$ population marginals, as illustrated in Figure~\ref{pd_corr_dists}.

\begin{figure}
\includegraphics[scale=0.6]{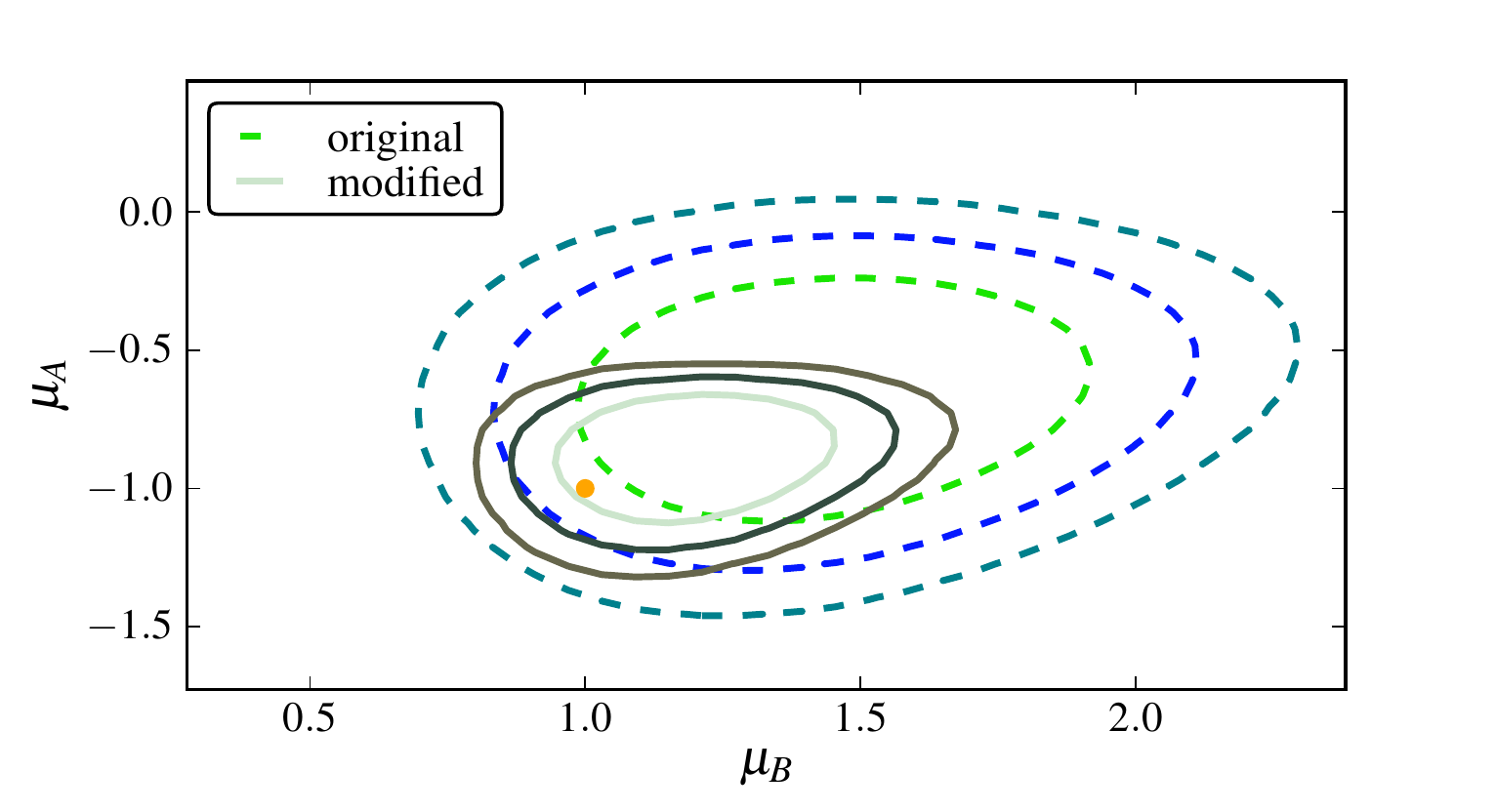} 
\caption{Posterior distributions on the parameters ($\mu_A$, $\mu_B$) using the correct posterior~\eqref{pp} (solid) and the \ori{} posterior~\eqref{BEAMSWE} (dashed). The \ori{} posterior assumes independence between $D|T$ and $P$.  Plotted are the 80\%, 95\% and 99\% confidence levels. The true parameters (orange point) lie within the 95 \% confidence regions of both posteriors.}
\label{post_conts}
\end{figure}

To compare the use of the \ori{} BEAMS posterior~\eqref{BEAMSIND} with the full conditional posterior~\eqref{BEAMSINDTRUE}, we randomly draw 40 data points from the above distribution and construct the respective posterior distributions, as illustrated in Figure~\ref{post_conts}. Observe that the \ori{} posterior is significantly wider than the full posterior. Indeed, approximately half of the interior of the 80\% region of the \ori{} posterior is ruled out to 1\% by the full posterior. It is interesting to note that, while the \ori{} posterior is wider than the full posterior, it is not biased. This result goes against our intuition; we believed that the \ori{} posterior would result in estimates for $\mu_A$ and $\mu_B$ which exaggerated $|\mu_A - \mu_B|$.  Whether it is a general result that no bias exists when the \ori{} posterior is used, or if there can exist dependencies between $P$ and $D$ for which the use of~\eqref{BEAMSWE} leads to a bias, remains an open question.

Figure~\ref{post_conts} illustrates one realistation from the distribution we have described, but repeated realisations show that on average, the variance in the maximum likelihood estimator using the \ori{} posterior is $\sim$\,3 times larger than the variance using the modified posterior. While these simulations are too simple to draw conclusions about cosmological parameter estimation from, they do suggest that where correlations between \btps{} and distance moduli exist within a class of SNe, it may be worthwhile accounting for it by using the modified posterior. Currently it is most common when modelling SNe for cosmology, to assume that the likelihood $f_{D|\Theta, T}(d | \theta, \tau)$ is a Gaussian with unknown mean and variance,
\begin{equation*}
D|\theta, P, T = \text{Normal}(\mu(\theta, T), \sigma(T)^2). 
\end{equation*}

If one wishes to include the \btps{} in the likelihood, one could include a linear shift in $P$ for the mean or variance. That is,

\begin{equation*}
D|\theta, P, T = \text{Normal}(\mu(\theta, T) + c_1 P, \sigma(T)^2 + c_2 P). 
\end{equation*}

Of course this is just one possibility, and one would need to analyse SN data to get a better idea of how $P$ should enter into the above equation.

\section{Obtaining Unbiased {\Large{$\tau_A$}}-probabilities}
\label{obtubp}

In this section we investigate likely sources of~\btp{} biases such as those presented in Section~\ref{basicbias}, and discuss how to detect and remove them. For SNe, one source of~\btp{} bias could be the failure to take into account the preferential confirmation of bright objects. This type of bias has been considered in the machine learning literature under the name of selection bias, and we here present the relevant ideas from there. We end the section with a brief discussion on how one could model the pdfs $f_{D | \Theta, \bs{P}, \bs{T}}$ and $f_{D | \Theta, \bs{F}, \bs{P}, \bs{T}}$, which are the likelihoods appearing in the extended posteriors introduced in Section~\ref{impa}.

\subsection{Selection Bias}
\label{selbias}
With respect to classification methods, selection bias refers to the situation where the confirmed data is a non-representative sample of the unconfirmed data. A selection bias is sometimes also referred to as a covariate shift although the two are defined slightly differently, as described in~\cite{Bickel07}. With selection bias, the confirmed data set is first randomly selected from the full set, and then at a second stage it is non-randomly reduced. Such is the situation with a population census, where at a first stage, a random sample of people is selected from the full population, and then at a second stage, people of a certain disposition cooperate more readily than others, resulting in a biased sample of respondees. 

A form of selection bias which is well known in observational astronomy is the Malmquist bias, whereby magnitude limited surveys lead to the preferential detection of intrinsically bright (low apparent magnitude) objects. In the case of SN cosmology, the bias is also towards the confirming of bright SNe. A reason for this bias is that the telescope time required to accurately classify a SN is inversely proportional to the SN's brightness. It is therefore relatively cheap to confirm bright objects and expensive to confirm faint ones. 

If the SN confirmation bias is ignored, certain inferences made about the global population of SNe are likely to be inaccurate. In particular, estimates of a classifier's False Positive and False Negative Rates will be biased, and the estimated \btps{} will be biased in certain circumstances, as we will discuss in the following section. 
\subsubsection{Formalism}

Following where possible the notation of~\cite{Fan05}, in what follows we assume that variables ($X$, $T$, $F$) are drawn from $\mathcal{X} \times  \mathcal{T} \times \mathcal{F}$, where

\begin{enumerate}
\item $\mathcal{X}$ is the feature space,
\item $\mathcal{T} = \{A, B \}$ is the binary type space,
\item $\mathcal{F} = \{0, 1 \}$ is the binary confirmation space, where $F = 1$ if \cfed{} ($F$ for \emph{f}ollowed-up).
\end{enumerate}

\noindent A realisation ($x$, $\tau$, $f$) lies in either the test set or the training sets, defined respectively as: 

\medskip 

\noindent \emph{test set} $  \stackrel{\text{def}}{=} \hspace{0mm} \{ (x, \tau, f) \hspace{3mm}\text{s.t.} \hspace{3mm} f = 0\}$

\noindent \emph{training set} $  \stackrel{\text{def}}{=} \hspace{0mm} \{ (x, \tau, f) \hspace{3mm}\text{s.t.} \hspace{3mm} f = 1\}$.

\medskip

\noindent For SN cosmology it could be that $\mathcal{X}, \mathcal{T}$ and $\mathcal{F}$ are respectively,

\begin{enumerate}
\item $\mathcal{X}$ is the space of all possible photometric data, where a SN's photometric data consists of apparent magnitudes and observational standard deviations in four colour bands over several nights.  
\item $\mathcal{T} = \{\text{Ia}, \text{nIa} \}$, type Ia and non-Ia SNe.
\item $\mathcal{F} = \{0, 1 \}$, where $F = 1$ if the SN has been spectroscopically \cfed{} and thus has its type known.
\end{enumerate}
\noindent By having a training set be \emph{unbiased} we mean that it is a representative sample of the test set, specifically that $F$ is independent of both $X$ and $T$. That is, the probability of confirmation is independent of features and type:
\begin{equation}
\label{kegarne}
\textrm{P}( F = 1 | X = x, T = \tau) = \textrm{P}(F = 1).
\end{equation} 

When the training set is unbiased, training set and test set objects are drawn from the same distribution over $\mathcal{X} \times \mathcal{T}$. This distribution over $\mathcal{X} \times \mathcal{T}$ can be estimated from the training set, so directly providing an estimate of the more useful test set distribution.

There are three important ways in which the independence relation~\eqref{kegarne} can break down, resulting in a \emph{biased} training set, as described in \cite{Zadrozny1} and listed below. By \emph{removing bias} from a training set, we mean reweighting the training points such that the training set becomes unbiased.

\begin{enumerate} 
\item  Confirmation is independent of features only when conditioned on type: $F |\, T$ and $X$ are independent. This is the simplest kind of biasing, and there are methods for correcting for it \citep{Bishop95}, \citep{Elkan01}. This is not the bias which exists in SN data.
\item Confirmation is independent of type only when given features: $F | X$ and $T$ are independent. If the decision to confirm is based on $X$ and perhaps some other factors which are independent of $T$, this is the bias which exists. This is probably the bias which exists in SN data, and there are methods for correcting for it, as we will discuss.\label{ii}
\item Confirmation depends on both features and type simultaneously. In this case, it is not possible to remove the bias from the data unless the exact form of the bias is known.  
\end{enumerate}
The decision to confirm a SN can be dictated by different features, examples include~\cite{Sako07, Sulli06}, all of which are contained in the photometric data $X$. Such was the also case in the SNPCC where the probability of confirmation was based entirely on the peak magnitude in the $r$ and $i$ f, as we will discuss in Section~\ref{snphotcc}. In reality, there are other factors which affect the confirmation decision such as the weather and telescope availability, but these are independent of SN type. Therefore the type~\ref{ii} bias above is the bias which exists in the SN data. Thus, for the remainder of this section we'll assume the type~\ref{ii} bias, that is
\begin{equation}
\label{as7}
\textrm{P}(F = 1 | X = x, T = \tau) = \textrm{P}(F = 1| X = x).
\end{equation}
The assumption of the type~\ref{ii} bias can be made stronger. The decision to confirm an object does not in general depend on all of $X$ but only a low-dimensional component $(X_F)$ of it, and so we have 
\begin{equation}
\label{as1}
\textrm{P}(F = 1| X = x, T = \tau) =  \textrm{P}(F=1 | X_F = x_F),
\end{equation}
where $X_F$ is contained in $X$. For SNe, $X_F$ could be the peak apparent magnitude in certain colour bands. 

In the following subsection we will describe how to correctly obtain \btps{} under the assumption of a bias described by~\eqref{as1}.
\subsection{Correctly obtaining \btps{}}
\label{obtBTP}
Let us remind the reader as to how we defined~\btps{} in the introduction:

\begin{equation}
\label{btpdef2}
\text{\btp{}} \stackrel{\text{def}}{=}  \textrm{P} (T_i = A | X_{P,i} = x_{P, i}) = p_i,
\end{equation}

where $X_{P,i}$ is an observable feature of the $i$th object, extracted from $X_i$. 
Estimates of $p_i$ values can be obtained using several methods, of which those mentioned previously are~\cite{Pozn1, newlingetal, INCA,2007AA46611G} It is worth rementioning that these different methods attempt to estimate different probability functions, as they each condition on different SN features. Thus there is no sense in which one set of~\btps{}  estimates is \textit{the} correct set.

We now make an adjustment to definition~\eqref{btpdef2}, to take into account that biased follow-up may result in an additional conditional dependence on $F$:

\begin{equation}
\label{btpdef}
\text{\btp{}} \stackrel{\text{def}}{=}  \textrm{P} (T_i = A | F_i = f_i, X_{P,i} = x_{P, i}) = P_i.
\end{equation}

The most informative \btps{} one could use would be those conditional on all of the features at one's disposal, 
\begin{equation}
\label{bopr}
X_P = X: \hspace{4mm} p_i = \textrm{P}(T = A | F = 0, X= x).
\end{equation}

However, when $\mathcal{X}$ is a high-dimensional non-homogeneous space, as is the case with photometric SN data, it can be difficult to approximate~\eqref{bopr} accurately. It is for this reason that it is necessary to reduce the features to a lower dimensional quantity $X_P \in \mathcal{X}_P$,  so that the \btps{} are calculated from a subspace ($\mathcal{X}_P$) of the full feature space, as described by~\eqref{btpdef}. The subspace $\mathcal{X}_P$ should be chosen to retain as much type specific information as possible while being of a sufficiently low dimension. In the SNPCC~\cite{newlingetal} chose $\mathcal{X}_P$ to be a 20-dimensional space of parameters obtained by fitting lightcurves. 

The job of obtaining estimated \btps{} for test set objects ($F = 0$) is one of obtaining an estimate of the type probability mass function, 
\begin{equation}
\label{desired}
f_{T | F, X_P}.
\end{equation}

\noindent Again, for~\eqref{desired} we prefer not to use the standard mass function notation, in order to to neaten certain integrals which follow. The \btp{} of a test set object can now be expressed in the following way, 
\begin{equation*}
\textrm{P}(T = A| F = 0, X_P = x_P) = f_{T | F, X_P}(A | 0, x_P).
\end{equation*}

Using kernel density estimation, boosting, or any other method of approximating a probability function, one can construct an approximation ($\hat{f}$) of the type probability function for training set objects,
\begin{equation}
\label{cadoo}
\hat{f}(x_P) \approx f_{T | F, X_P}(A | 1, x_P).
\end{equation}

Using the estimate $\hat{f}$ in~\eqref{cadoo} one can estimate the~\btps{} for the training set objects:
\begin{equation}
\label{pere}
\textrm{P}(T = A | F = 1, X_P = x_P) \approx \hat{f}(x_P).
\end{equation}

The estimate~\eqref{pere} is not directly important as the training set object types are known exactly. But it is only through the training set objects that we can learn anything about the types of the test set objects. 

How $\hat{f}$ from the training set is related to $f_{T|F = 0,X_P}$~\eqref{desired} depends on the relationship between $X_F$ (the data which determines confirmation probability) and $X_P$ (the data used to calculate \btps{}). There are two cases to consider. The first, which we write as $\mathcal{X}_F \subset \mathcal{X}_P$, is when the data which determines confirmation probabilities is completely contained in the data used to calculate \btps{}. That is, 
\begin{equation}
\notag
X_F \subset X_P\hspace{4mm} \stackrel{\text{def}}{\leftrightarrow}\hspace{4mm} \textrm{P}(F = 1 | X_P = x_P) = \textrm{P}(F = 1 |X_F = x_F).
\end{equation}

The second case, when $\mathcal{X}_F \not \subset \mathcal{X}_P$ is when not all confirmation information is contained in $X_P$,   
\begin{equation}
\notag
\mathcal{X}_F \not \subset \mathcal{X}_P\hspace{4mm} \stackrel{\text{def}}{\leftrightarrow}\hspace{4mm} \textrm{P}(F = 1| X_P = x_P) \not = \textrm{P}(F =1 |X_F = x_F).
\end{equation}

\noindent In the case of $\mathcal{X}_F \subset \mathcal{X}_P$, it can be shown that, 
\begin{equation}
\label{useful1}
\textrm{P}(F = 1|T = \tau, X_P = x_P) = \textrm{P}(F = 1 | X_F = x_F).
\end{equation}

\subsubsection{$\mathcal{X}_F \subset \mathcal{X}_P$}
\label{subsoosa}

We will show that in the case of $\mathcal{X}_F \subset \mathcal{X}_P$, a type probability function approximating the training population $(\hat{f})$ is an unbiased approximation for the type probability function of the test population $(F = 0)$. To show this we start with the type probability of a test object:
\begin{align}
\textrm{P}(T& = \tau | F = 0, X_P = x_P).  \notag \\
\intertext{$\rightarrow$ Using Bayes' Theorem, we have} \notag
& = \frac{\textrm{P}(F = 0| T= \tau, X_P = x_P) \cdot \textrm{P}(T= \tau | X_P = x_P)}{\textrm{P}(F = 0 | X_P = x_P)} \notag\\
\intertext{$\rightarrow$Then using~\eqref{useful1}, we have} \notag
& = \frac{\textrm{P}(F = 0|X_F = x_F) \cdot \textrm{P}(T = \tau | X_P = x_P)}{\textrm{P}(F = 0 | X_F = x_F)} \notag \\
& = \textrm{P}(T = \tau | X_P = x_P). \label{post_comments} \\
\intertext{$\rightarrow$ Using the same steps as above but in reverse and with $F = 1$, we arrive at} \notag
& = \textrm{P}(T = \tau | F = 1, X_P = x_P). \notag \\
\intertext{$\rightarrow$ This is the type probability function for training set objects, and it can be approximated:} 
& \approx  \hat{f}(x_P). \label{hand1}
\end{align}

This is a useful result, as it says that $\hat{f}$ is not only an approximation of the type probability function of the training data, but also of the test set. Thus, $\hat{f}$ should provide unbiased \btps{} for the test set when $\mathcal{X}_F \subset \mathcal{X}_P$.

It should be noted that for $\hat{f}$ to be a good approximation for the test set, it is necessary that the training set covers all regions of $\mathcal{X}_P$ where there are test points. That is, if there are values of $x_P$ for which $\textrm{P}(X_P = x_P|F=1) = 0$ and $\textrm{P}(X_P = x_P|F=0) \not= 0$, then the approximation $\hat{f}$ will not converge to $f_{T|F = 0,X_P}$ as the training set size grows. One can refer to~\citet{Fan05} for a full treatment of this topic. 

With respect to SNe, the requirement of the preceding paragraph is that, if a SN is too faint to be confirmed and to enter the training set, it should not enter the test set either. We will return to this point again in Section~\ref{snphotcc}. 

One important question which we do not attempt to answer here is, how many SNe of different apparent magnitudes should be \cfed{} to obtain as rapid as possible convergence of $\hat{f}$ to $f_{T|F = 0,X_P}$. An interesting method for deciding which SNe to confirm may be one based on the real-time approach proposed in~\cite{Freund97}, where the decision to add an object to the training set is based on the uncertainty of its type using the currently fitted $\hat{f}$. In Section~\ref{snphotcc} we discuss this further.

\subsubsection{$\mathcal{X}_F \not \subset \mathcal{X}_P$}
If $\mathcal{X}_F \not \subset \mathcal{X}_P$ we will not be able to use $\hat{f}$ to estimate the~\btps{} in the test set, as~\eqref{hand1} required $\mathcal{X}_F \subset \mathcal{X}_P$. In addition to this problem of not being able to use $\hat{f}$ to obtain unbiased \btps{} for the test set objects, if $\mathcal{X}_F \not \subset \mathcal{X}_P$ then
\begin{equation*}
\textrm{P}(T = \tau | X_P = x_P) \not = \textrm{P}(T | X_P = x_P, X_F = x_F).
\end{equation*} 
This tells us that there is additional type information to be obtained from $X_F$, and so by not including $X_F$ one is wasting type information. For this reason we recommend reconstructing the \btps{} based on redefined features, $X_P \leftarrow (X_F, X_P)$. 

However, it is possible that one explicitly does not want to use $X_F$ in calculating \btps{}. This may be the case if one wishes to reduce the dependence between $D$ and $P$, as presented in Section~\ref{impa}. For SNe, this may involve obtaining \btps{} from shape alone, independent of magnitude, so that $\mathcal{X}_P$ is a space whose dimensions describe only shape and not magnitude. In this case, as we cannot use $\hat{f}$, we need to use the relationship derived in~\cite{Shimo},

\begin{align}
& \hspace{0mm} \textrm{P}(T = \tau | F = 0, X_P)  \notag \\
& = \int_{\mathcal{X}_F} f_{T, X_F | F, X_P} (\tau, x_F | 1, x_P) \cdot w(x_F,x_P) \, dx_F  \label{weight22},
\end{align}

where the weight function is defined as

\begin{equation}
\label{lew}
w(x_F, x_P) = \frac{f_{F|X_F}(0| x_F)f_{F|X_P}(1| x_P)}{f_{F|X_F}(1| x_F)f_{F|X_P}(0| x_P)}.
\end{equation}

Notice that if $\mathcal{X}_F \subset \mathcal{X}_P$, then $w(x_F, x_P) = 1$ and so~\eqref{weight22} reduces to the type probability function for training set objects, approximated by $\hat{f}$ as expected from~\eqref{hand1}. When $w(x_F, x_P) \not= 1$, the training set type probability function $\hat{f}$ cannot be used directly as an approximation to the test set type probability function. However, if each training set object is weighted using~\eqref{weight22}, then an unbiased test set type probability function approximation can be obtained.

The weight function~\eqref{lew} does not require any type information and so can be estimated as a first step. This additional step of estimation introduces additional error into the final estimate of~\eqref{desired}, a theoretical analysis of which is presented in~\cite{bias_correction_error}. An alternative to the two-stage approach would be to fit the two terms in~\eqref{weight22} simultaneously, as suggested and described by~\citep{Bickel07}. The use of~\eqref{lew} was first suggested in~\cite{Shimo}, where a detailed analysis of the asymptotic behaviour of its approximation is given. Therein, it is suggested that~\eqref{lew} be approximated by kernel density estimation.

In the case where $F$ and $X_P$ are independent, the weight function reduces to one of only $X_F$, 
\begin{align}
\label{weight_ind}
w(X_F = x_F) &= \frac{\textrm{P}(F = 0|X_F = x_F)\textrm{P}(F = 1)}{\textrm{P}(F = 1|X_F = x_F)\textrm{P}(F = 0)}.
\end{align}

This reduction in dimension may be valuable in approximating the weight function. 

\subsection{Detecting and removing biases in \btps{}}
In the previous section we presented the correct way in which to estimate \btps{} in the case $\mathcal{X}_F \not\subset \mathcal{X}_P$. In this section we will present an example illustrating this process, but in the context of bias removal. 

Suppose that we have a program which outputs scalar values ($\tilde{p}$), which are purported \btps{}. We believe that the output values have some unspecified bias, which we wish to remove. An assumption we make is that the $\tilde{p}$ values are calculated in the same way for training and test sets. That is that the program does not process cases $F=0$ and $F=1$ differently. It may seem strange to be interested in what the program does when $F=1$, but as already mentioned it is only from the training set that we can learn anything about the test set. The idea now is to treat the received $\tilde{p}$ values as the $x_P$s from the previous section, and not directly as \btps. 
 
For this example, we choose $\mathcal{X}_F = [0,1]$. To now transform a test set value $\tilde{p} \in [0,1]$ into an unbiased \btp{} using~\eqref{weight22}, one needs to estimate certain probability functions using kernel density estimation. The necessary functions we see from~~\eqref{weight22} and \eqref{lew} are $f_{T, X_F | F, X_P}(\tau, x_F | 1, \tilde{p}$), $f_{F | X_F}(1, x_F)$, $f_{F|X_P}(0, \tilde{p})$,  $f_{F|X_F}(0, x_F)$ and $f_{F | X_P}(0, x_P)$.

It is an interesting and important question as to how accurately these probability functions can be approximated with few data points, but for this example we assume them known,

\begin{align}
& f_{T, X_F | F, X_P}(A, x_F | 1, \tilde{p})  = \left\{ \begin{array}{lll} 
x_F & \mbox{if  } \frac{1}{2} x_F^2 < \tilde{p} < 1 - \frac{1}{2} x_F^2, \notag \\
2 \cdot x_F & \mbox{if  } \tilde{p} > 1 - \frac{1}{2} x_F^2 \notag \\
0  & \mbox{if  } \tilde{p} < \frac{1}{2} x_F^2. 
\end{array} \right. \notag \\
&f_{F|X_F}(0, x_F)  = (1 - x_F), \notag \\
&f_{F | X_F}(1, x_F)  = x_F, \label{exogen} \\
& f_{F|X_P}(1|\tilde{p}) =  f_{F|X_P}(0|\tilde{p})  = \frac{1}{2}. \notag
\end{align} 

\begin{figure}
\includegraphics[scale=0.6]{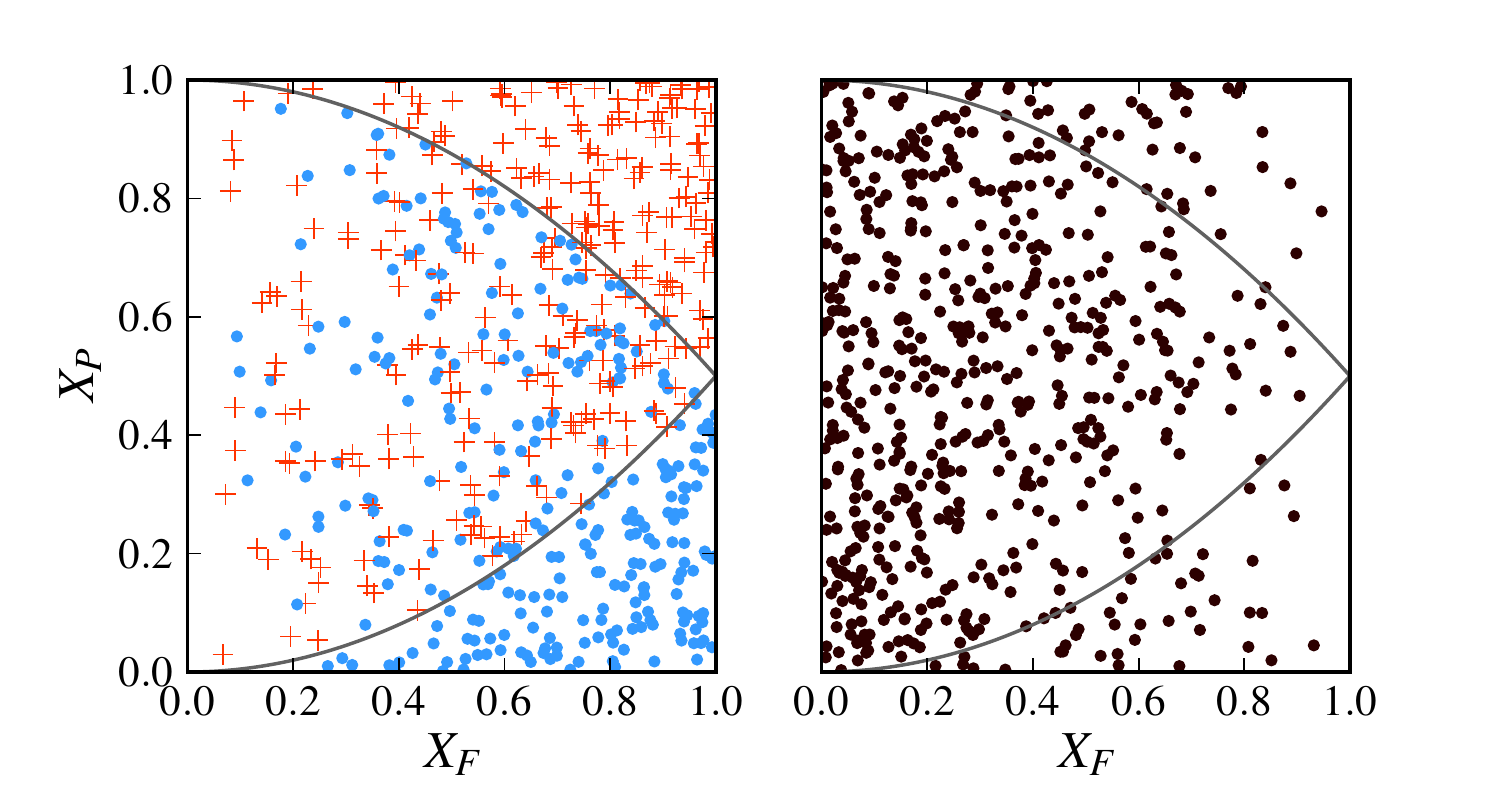} 
\caption{Realisations of a training set (left) containing type A (red pluses) and type B (blue points) objects, and a test set (right), drawn according to~\eqref{exogen}. Overlaid are faint lines delineating the discrete regions described by~\eqref{exogen}}
\label{bia_wei}
\end{figure} 
Realisations from the above distribution are illustrated in Figure~\ref{bia_wei}. By integrating $x_F$ out of $f_{T, X_F | F, X_P}(A, x_F | 1, \tilde{p})$ in~\eqref{exogen}, we have that 
\begin{equation}
\label{Fe1t}
\textrm{P}(T = A | F = 1, X_P = \tilde{p}) = \tilde{p}.
\end{equation}
That is, in the training set $\tilde{p}$ is an unbiased estimate of a \btp{}. The \btps{} for objects in the test set we estimate using~\eqref{weight22},

\begin{align}
\textrm{P}(T = A | &F = 0, X_P = \tilde{p}) \notag \\
& = \int_{\mathcal{X}_F} f_{T, X_F|F, X_P} (\tau, x_F | 0, x_P) \, dx_F \hspace{20mm} \notag \\
& = \int_{\mathcal{X}_F} f_{T, X_F|F, X_P} (\tau, x_F | 1, x_P)\,  w(x_F, \tilde{p}) \, dx_F \notag \\ 
& = \int_{\mathcal{X}_F} f_{T, X_F|F, X_P} (\tau, x_F | 1, x_P)\,  \frac{1 - x_F}{x_F} \, dx_F \notag\\ 
& =  \left\{ \begin{array}{ll} 
\sqrt{2 \,\tilde{p}} - \tilde{p} & \mbox{if  } \tilde{p} < 0.5,  \label{postcprr} \\
2 - \tilde{p} - \sqrt{2 - 2 \tilde{p}} & \mbox{if  }  0.5 < \tilde{p}.   
\end{array} \right.
\end{align}

\begin{figure}
\includegraphics[scale=0.6]{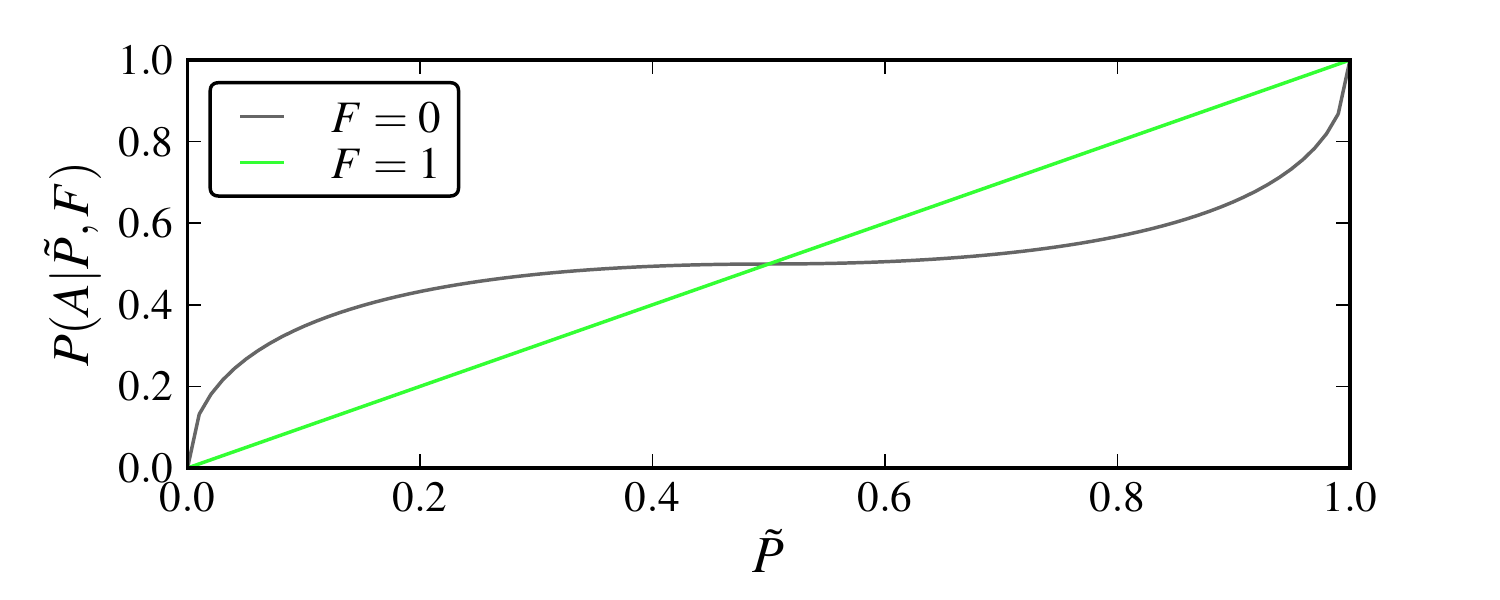} 
\caption{Corrected \btps{}. The disproportionately large number of training SNe with decisive \btps{} (as depicted in Figure~\ref{bia_wei}), causes $\tilde{p}$ values to be too confident as test set \btp{}  estimates.}
\label{correctd}
\end{figure}

The \btps{}~\eqref{Fe1t} and~\eqref{postcprr} are plotted in Figure~\ref{correctd}, where we see that $\tilde{p}$ provided accurate \btps{} for the training set, but not for the test set. This is not unexpected in reality, where the program providing the \btps{} may have been trained only on the biased training data. It is important to remember that this bias should only arise when $\mathcal{X}_F \not \subset \mathcal{X}_P$.

\section{Supernova surveys and the SNPCC}
\label{snphotcc}

The SNPCC provided a simulated spectroscopic training data set of approximately 1000 known SNe. The challenge was then to predict the types of approximately $20\,000$ other objects\footnote{These lightcurves are available at http://sdssdp62.fnal.gov/ sdsssn/SIMGEN\_PUBLIC/} from their lightcurves alone. Since the end of the competition, the types of all the simulated SNe have been released, making a post competition autopsy relatively easy to perform. In the results paper~\cite{sntc_results} we see that the probability that a SN was \cfed{} was based on the $r$-band and $i$-band quantities,  

\begin{align*}
\epsilon_{\text{spec}}^{\text{band}} = \epsilon_0\left(1 - x^{l}\right) & \hspace{7mm} x \stackrel{\text{def}}{=}  
\frac{ m_{\text{peak}}^{\text{band}} - M_{\text{min}}^{\text{band}} }{ m_{\text{lim}}^{\text{band}} -  M_{\text{min}}^{\text{band}}}. 
\end{align*}

where $m^{\text{band}}_{\text{peak}}$ is the band-specific apparent magnitude of a SN, and $M^{\text{band}}_{\text{min}}$ and $m^{\text{band}}_{\text{min}}$ are constants. In~\cite{sntc_results} it is given that for col $=r$ and col $=i$,

\begin{align}
\label{constiti}
\epsilon_{\text{spec}}^{r} = \epsilon_0\left(1 - x^5\right) & \hspace{7mm} x  \stackrel{\text{def}}{=} \frac{m^{r}_{\text{peak}} - 16.0}{5.5}  \\
\epsilon_{\text{spec}}^{i} = \epsilon_0\left(1 - x^6\right) & \hspace{7mm} x  \stackrel{\text{def}}{=} \frac{m^{i}_{\text{peak}} - 21.5}{2.0}\notag
\end{align}
where $\epsilon_0$ is some constant. Once $\epsilon_{\text{spec}}^{i}$ and $\epsilon_{\text{spec}}^{r}$ have been calculated, if a $[0 \rightarrow 1]$ uniform random number is less than either of them, confirmation is performed. 
As confirmation depends only on $\epsilon_{\text{spec}}^{i}$ and $\epsilon_{\text{spec}}^{r}$, we have from~\eqref{post_comments} that
\begin{figure}
\includegraphics[scale = 0.58]{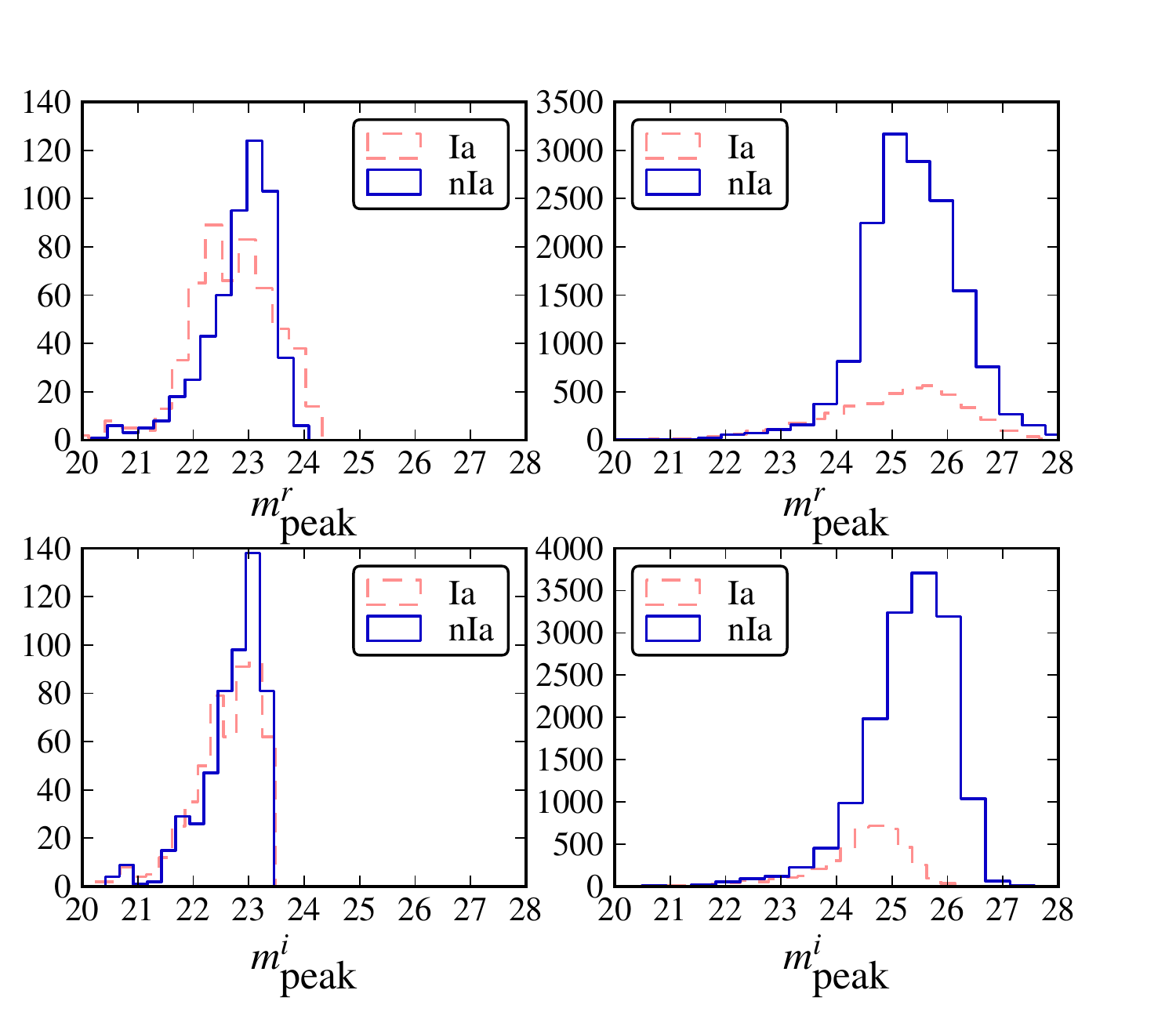} 
\caption{Counts of \cfed{} (left) and not \cfed{} (right) SNe, Ia (dashed) and non-Ia (solid) as a function of $m^{r}_{\text{peak}}$ (above) and $m^{i}_{\text{peak}}$ (below). }
\label{nir}
\end{figure}  
\begin{equation}
\label{lablab}
\textrm{P}(T = \tau | F = 0, m^{i}_{\text{peak}}, m^{r}_{\text{peak}}) = \textrm{P}(T = \tau | F = 1, m^{i}_{\text{peak}}, m^{r}_{\text{peak}}).
\end{equation} 
Equation~\eqref{lablab} can be interpreted as saying that the ratio Ia:nIa is the same in a given $m^{i}_{\text{peak}}, m^{r}_{\text{peak}}$ bin. 
The manner in which the follow-up was simulated should of course guarantee that~\eqref{lablab} holds. In theory one should be able to deduce the verity of~\eqref{lablab} from Figure~\ref{nir}, but the redshift bins with large numbers of confirmed SNe are too sparsely populated by unconfirmed SNe to check that the Ia:nIa is invariant. To be in a position where~\eqref{lablab} can be checked is in general an unrealistic luxury, as without the types of the test objects this is impossible.

In terms of obtaining accurate \btps{}, a disturbing feature of Figure~\ref{nir} is the absence of training SNe with high apparent magnitudes. With no training SNe with $i$-band apparent magnitudes greater than 23.5, we cannot infer the types of test SNe with apparent magnitudes greater than 23.5. Indeed there would be no non-astrophysical reason not to believe that all SNe with apparent magnitudes greater than 23.5 are non-Ia. As already mentioned in Section~\ref{selbias}, in situations where the training set does not span the test set, one should ignore unrepresented test objects from all analyses. All test SNe other than those for which there are training SNe of comparable peak apparent magnitudes in $r$ and $i$ bands should be removed from a BEAMS analysis, unless there is a valid astrophysical reason not to do so. This entails ignoring about 95\% of unconfirmed SNe; an enormous cut. We therefore consider it important to confirm more faint SNe.

In~\cite{newlingetal}, a comparison is made between training a boosting algorithm on the non-representative spectroscopically \cfed{} SNe and a representative sample, randomly selected from the unconfirmed SN set. Therein, the authors use twenty fitted lightcurve parameters, including fitted apparent magnitudes in $r$ and $i$ bands. This corresponds to the situation discussed in Section~\ref{subsoosa}, where $\mathcal{X}_P \subset \mathcal{X}_F$. For this reason, the probability density function $\hat{f}$ in~\ref{pere} as estimated by their boosting algorithm should be an unbiased estimate for $f_{T|F = 0,X_P}$. But being unbiased does not guarantee low error, and when trained on the \cfed{} SNe, regions of parameter space corresponding to high apparent magnitude had no training SNe with which to learn, and so the approximation of~\ref{desired} was poor. However when trained on the representative set, every region of populated parameter space was represented by the training set, and the approximation of~\ref{desired} was greatly improved.

In their paper,~\cite{INCA} describe their entry in the SNPCC, and they report how a semi-supervised learning algorithm performs better with a few faint training SNe than with many bright ones. The comparison was performed while keeping the total confirmation time constant. Thus their conclusion was the same as ours; that it is important to obtain a more representative SN training sample.

\section{Conclusions and Recommendations}
\label{conclusions}

In this paper we discussed BEAMS, and 
extended the \ori{} posterior probability function to the case when $D|T$ (distance modulus $|$ type) and $P$ (type probability) are dependent. In Section~\ref{dpcorr} we considered an example where the dependence between $D|T$ and $P$ is strong, and observed a large reduction in the posterior width using the extended posterior as opposed to the \ori{} posterior. No bias is observed when using either the extended or the \ori{} posterior.

In Section~\ref{sec:sims} we considered examples where the \ori{} posterior is valid, that is when $D|T$ and $P$ are independent. We performed tests to ascertain the importance to BEAMS of i) the \emph{decisiveness} of the \btps{} (observations of $P$), and ii) \emph{sample size}. In one test~\eqref{sim2}, we observed how doubling a sample size reduces error in parameter estimation more than obtaining the true type identity of the objects does. In another test~\eqref{sim22}, we observed how BEAMS accurately locates two population means, but fails to match each mean to its population.

We looked at the effects of using biased \btps{} in Section~\eqref{basicbias}. The result of~\KBH{}, that \btp{} biases towards population $A$ affect the population's parameter estimates less than biases in favour of population $B$, was observed. A similar result which is uncovered is that biases towards high decisiveness are more damaging than biases towards low decisiveness. In other words, it is better to be conservative in your prior type beliefs than too confident.

Our recommendations for BEAMS may thus be summarised as follows. Firstly, the inclusion in the likelihood function of \btps can dramatically reduce the width of the final posterior, providing tighter constraints on cosmological parameters. Secondly, conservative estimation of \btps{} is less harmful than too decisive an estimation. Thirdly, it is possible to remove biases in \btps{} using the techniques described in Section~\eqref{obtubp}.

In Section~\ref{obtubp} we considered the problem of debiasing \btps{}. Interpreting recent results from the machine learning literature in terms of SN cosmology, we discussed the different ways in which training sets can be biased and how to remove such biases. The key to understanding and correcting biases is the relationship between $\mathcal{X}_F$ and $\mathcal{X}_P$, where $\mathcal{X}_F$ are object features which determine \emph{confirmation probability}, and $\mathcal{X}_P$ are those features which determine \emph{\btps{}}. In brief, when $\mathcal{X}_P$ contains $\mathcal{X}_F$, \btps{} should be unbiased, but if this is not the case, there are sometimes ways for correcting the bias.

With respect to future SN surveys, we emphasize the importance of an accurate record as to what information is used when deciding whether or not a SN is \cfed{}. Using this information, one should in theory be able to remove all the affects of selection bias when $\mathcal{X}_F \not \subset \mathcal{X}_P$. In other words, using all the variables which are considered in deciding whether to follow-up a SN, it will always be possible to obtain unbiased \btps{}, irrespective of what the \btps{} are based on. Such follow-up variables may include early segments of light curves, $\chi^2$ goodness of fits, fit probabilities, host galaxy position and type, expected peak apparent magnitude in certain filters, etc.
 
Our second recommendation for SN surveys is that more faint objects are \cfed{}. While it not necessary for most machine learning algorithms to have a spectroscopic training set which is exactly representative of the photometric test set, it is necessary that the spectroscopic set at least covers the photometric set. Thus having large numbers of faint unconfirmed objects without any confirmed faint objects is suboptimal.

\section{Acknowledgements}
JN has a SKA bursary and MS is funded by a SKA fellowship. BB acknowledges funding from the NRF and Royal Society. MK acknowledges financial support by the Swiss NSF. RH acknowledges funding from the Rhodes Trust.

\appendix
\section{Posterior Type Probabilities}

We here derive the posterior type probabilities based on the modifications of Section~\ref{impa}. The posterior type probability will be derived, conditional on $\boldsymbol{D}$ and $\boldsymbol{P}$. This derivation can be easily extended to posterior type probabilities conditional on $\boldsymbol{D}, \boldsymbol{F}$ and $\boldsymbol{P}$.

\begin{align}
& f_{T_i | \bs{D}, \bs{P}}(A | \bs{d}, \bs{p}) \nonumber \\
&= \int_{\theta}f_{T_i | \Theta, \bs{D}, \bs{P}}(A | \theta, \bs{d}, \bs{p}) f_{\Theta|\bs{D}, \bs{P}}(\theta | \bs{d}, \bs{p}) \,d\theta \nonumber \\
&= \int_{\theta}f_{T_i | \Theta, D_i, P_i}(A | \theta, d_i, p_i) f_{\Theta|\bs{D}, \bs{P}}(\theta | \bs{d}, \bs{p}) \,d\theta \nonumber \\
\intertext{$\rightarrow$ we have assumed that the objects are independent,}
&= \int_{\theta} \frac{f_{D_i|\Theta, P_i, T_i}(d_i |\theta, p_i, A)f_{T_i|\Theta, P_i}( A | \theta, p_i)}{f_{D_i|\Theta, P_i}(d_i | \theta, p_i)}  \hspace{2mm} \times \nonumber \\
& \hspace{50mm} \times \hspace{2mm} f_{\Theta | \bs{D}, \bs{P}} (\theta | \bs{d}, \bs{p}) \, d\theta \nonumber \\ 
\intertext{$\rightarrow$ we have used Bayes' Theorem,}
&= \int_{\theta} \left(\frac{A_i}{A_i + B_i}\right) f_{\Theta | \bs{D}, \bs{P}}(\theta | \bs{d}, \bs{p})   \, d\theta. \label{poster} 
\intertext{$\rightarrow$ where $A_i = \textrm{P}(d_i |\theta, p_i, T_i = A)p_i$, $ B_i = \textrm{P}(d_i |\theta, p_i, $  $ T_i = B)(1-p_i),$ and we have assumed used that $f_{T_i|\Theta, P_i}( A | \theta, p_i) = p_i$.}\notag
\end{align}

 If the posterior $f_{\Theta|\bs{D}, \bs{P}}$  confines $\theta$ to a region sufficiently small such that $A_i$ and $B_i$ are approximately constant, then the posterior type probability~\eqref{poster} is well approximated by $A_i(\hat{\theta}) / \left(A_i(\hat{\theta}) + B_i(\hat{\theta})\right)$ where $\hat{\theta}$ is the maximum likelihood estimator of $f_{\Theta | \bs{D}, \bs{P}} (\theta | \boldsymbol{d}, \boldsymbol{p})$. Furthermore, the posterior odds ratio,

\begin{equation*}
\text{posterior odds ratio } \stackrel{\text{def}}{=} \,\, \frac{f_{T_i | \bs{D}, \bs{P}}(A | \boldsymbol{d}, \boldsymbol{p})}{f_{T_i | \bs{D}, \bs{P}}(B | \boldsymbol{d}, \boldsymbol{p})}
\end{equation*}
can be shown to be given by the prior odds ratio multiplied by the Bayes Factor,
\begin{equation*}
\text{posterior odds ratio} = \left(\frac{p_i}{1-p_i} \right) \times \left(\frac{f_{D_i|\Theta, P_i, T_i}(d_i | \hat{\theta}, p_i, A)}{f_{D_i|\Theta, P_i, T_i}(d_i | \hat{\theta}, p_i, B)}\right).
\end{equation*}

\section{Additional conditioning on the confirmation of supernova type}
In this paper we did not distinguished between the contributions of unconfirmed and confirmed objects to the posterior. While we can calculate approximate \btps{} for confirmed objects, these values should not enter the posterior, but be replaced by 0 (if type B) or 1 (if type A). Let us introduce the random variable $F$ to denote whether an object is confirmed, so that $F = 1$ if confirmed and $F = 0$ if unconfirmed. With this introduced, we wish to replace the \btps{} $\bs{p}$ by $\bar{\bs{p}}$, where,
\begin{equation*}
\bar{p}_i =  \left\{ \begin{array}{lll} 	
p_i & \mbox{if } f_i = 0, \\
1 & \mbox{if } f_i = 1 \mbox{ and } \tau_i = A,\\
0 & \mbox{if } f_i = 1 \mbox{ and } \tau_i = B.
\end{array} \right.
\end{equation*} 
We must be careful to let the new information which we introduce in $\bar{\bs{p}}$ be absorbed elsewhere in the posterior. To this end, as we did in Section~\ref{dp} we start afresh the posterior derivation, explicitly including the vector ($\bs{f}$) which describes which objects have been followed-up. Doing this, we arrive at the following posterior distribution

\begin{align}
f_{\Theta | \bs{D}, \bs{F}, \bs{P}}(\theta |\bs{d}, \bs{f}, \bs{p})  \propto  & \, f_{\Theta}(\theta)  \hspace{2mm} \times \label{pp2} \\
\sum_{\boldsymbol{\tau}}  \, f_{\bs{D} |\Theta, \bs{F} ,\bs{P} , \bs{T} }&(\bs{d}|\theta,\bs{f}, \bs{p}, \bs{\tau})  \prod_{\tau_i = A}\bar{p}_i\prod_{\tau_j = B}(1 - \bar{p}_j). \notag
\end{align}

The new information ($\bs{f}$) has been absorbed into the likelihood, $f_{\bs{D}|\cdots}$. For a particular application, one may now ask if the addition of $\bs{F}$ in $f_{\bs{D}|\cdots}$ is necessary. We have already mentioned that for SNe $\bs{D}|\theta, \bs{T}$  is unlikely to be independent of $\bs{P}$. It is also unlikely that $\bs{D}|\theta, \bs{T}$ is independent of $\bs{F}$, as bright SNe, which have lower fitted distance moduli at a given redshift, are \cfed{} more regularly than faint ones. However, it is possible that by additionally conditioning $\bs{D}$ on $\bs{P}$ this confirmation dependence is broken, so that $\bs{D}|\theta,\bs{P},\bs{T}$ and $\bs{F}$ are independent. We leave this as an open question. 

In the case of independent SNe, the posterior~\eqref{pp2} reduces to
\begin{align}
f_{\Theta | \bs{D} , \bs{F} , \bs{P}}(\theta|\bs{d}, \bs{f},\bs{p}) \propto \prod_{i = 1}^{N} \big[ \, f_{D_i|\Theta, F_i, P_i, T_i}  \left(d_i |\theta, f_i, p_i, A\right) & \bar{p}_i \hspace{1.5mm} +   \notag  \\
f_{D_i|\Theta ,  F_i, P_i, T_i}(d_i | \theta, f_i, p_i, B) \, \left(1-  \bar{p}_i\right)  \big].& 
\label{BEAMSINDFTRUE}
\end{align}

\bibliographystyle{mn2e}
\bibliography{BEAMS_experiments1}{}

\end{document}